\documentclass[aps,prl,twocolumn,superscriptaddress]{revtex4}
\usepackage{mathrsfs}
\usepackage{epsfig}
\usepackage{graphicx}
\usepackage{amsfonts}
\usepackage[figuresright]{rotating}
\usepackage{amssymb}
\usepackage{amsmath}
\usepackage{dcolumn}
\usepackage{bm}
\usepackage{color}

\def\be{\begin{equation}} \def\ee{\end{equation}}
\def\bea{\begin{eqnarray}} \def\eea{\end{eqnarray}}

\newcommand{\WQCASQC} {Wilczek Quantum Center and Key Laboratory of Artificial Structures and Quantum Control, School of Physics and Astronomy, Shanghai Jiao Tong University, Shanghai 200240, China}

\newcommand{\SRCQC}{Shanghai Research Center for Quantum Sciences, Shanghai 201315, China}

\begin{document}
\title{Space-time symmetry breaking in nonequilibrium frustrated magnetism}

\author{Mingxi Yue}
\affiliation{\WQCASQC}

\author{Zi Cai}
\email{zcai@sjtu.edu.cn}
\affiliation{\WQCASQC}
\affiliation{\SRCQC}

\begin{abstract}  
Spontaneous symmetry breaking is responsible for the rich phenomena in equilibrium physics. Driving a system out-of-equilibrium  can significantly enrich the possibility of spontaneous symmetry breaking, which occurs not only in space, but also in time domain. This study investigates a driven-dissipative frustrated magnetic system. Results show   that  frustration in such a far-from-equilibrium system could lead to a wealth of intriguing non-equilibrium phases with intertwined space-time symmetry breaking, {\it e.g.}, a discrete time crystal phase accompanied by a time-dependent spatial order oscillating between a long-range tripartite stripe and a short-range ferromagnetic order.



\end{abstract}


\maketitle

{\it Introduction --} Frustration arises  when  interacting energies cannot be simultaneously minimized for all bonds in a many-body system. It hosts remarkable phenomena ranging from classical spin glass\cite{Mezard1986} to quantum spin liquid\cite{Lacroix2011}. Typically, frustration  may lead to macroscopic degeneracy in the classical ground-state manifold. However, this degeneracy could be  lifted by thermal or quantum  fluctuations, which may select particular configurations out of the degenerate manifold\cite{Villain1979,Henley1987,Chubukov1992}, or make a superposition among them to form exotic quantum states without classical counterparts\cite{Anderson1973}. For the past decades, frustrated systems have been one of the central themes in condensed matter physics. However, with a few  exceptions\cite{Wan2017,Wan2018,Bittner2020,Jin2022}, most studies have restricted their the scope within equilibrium physics, which is governed by the paradigm of  (free) energy minimization. The effect of frustration on nonequilibrium systems is far from clear, because these systems can absorb energy from external driving or environment, thus they are usually far from the ground state and the energy minimization principle does not necessarily apply.

 Compared to equilibrium physics, non-equilibrium physics is much richer, albeit less known. A  prototypical example is the paradigm of spontaneous symmetry breaking(SSB) that plays a crucial role in both equilibrium and nonequilibrium physics.  In contrast to SSB within the thermal equilibrium, which is rooted in the variational principle of (free) energy minimization, the richness of the spatiotemporal structures spontaneously emerging from far-from-equilibrium systems can only be understood within a dynamical framework, even for a nonequilibrium steady state\cite{Cross2009}.  The nonequilibrium phases of matter fundamentally differ from the equilibrium ones in the sense that the time dimension plays an equally, if not more, important role than the spatial dimensions in the classification of phases of matter. For instance, incorporating the time direction significantly enriches the possibility of SSB, giving rise to a wealth of intriguing  nonequilibrium phases, like the  time crystal that  spontaneously breaks the discrete or continuous time translational symmetry\cite{Wilczek2012,Bruno2013,Watanabe2015,Sacha2015,Else2016,Khemani2016,Yao2017,Choi2017,Zhang2017,Autti2018,Trager2021,Stehouwer2021,Mi2022,Frey2022,Kongkhambut2022}. Frustration is a source of the  exotic phase in equilibrium physics, thus, one may wonder whether it could lead to novel phases of matter with an intriguing space-time structure in far-from-equilibrium systems.

In this study, we attempt to address this question, by focusing on a driven-dissipative interacting spin model. Periodic driving usually heats  a generic closed many-body system toward an infinite temperature state. To avoid this indefinite energy absorption, we introduce dissipation by coupling each spin to a heat bath, which will drive the spin system to thermal equilibrium in the absence of periodic driving\cite{Supplementary}.  Considering the notorious difficulty of dealing with quantum many-body dynamics, we focus on a classical system that enables us to simulate high-dimensional systems up to a large system size.   It has been shown that the exotic nonequilibrium phases are not restricted to quantum systems ({\it e.g.} the time crystal phases in classical many-body systems have recently been investigated\cite{Lupo2019,Yao2020,Gambetta2019,Khasseh2019,Hurtado2020,Ye2021,Pizzi2021}). Different from previous studies about the prethermal dynamics of close systems\cite{Jin2022,Ye2021,Pizzi2021}, here we focus on the long-time asymptotic behavior of the driven-dissipative system, and show that incorporating frustration significantly enriches the categories of  nonequilibrium phases of matter, and leads to intriguing magnetic states with intertwined space-time symmetry breaking.



 \begin{figure}[htb]
\includegraphics[width=0.8\linewidth]{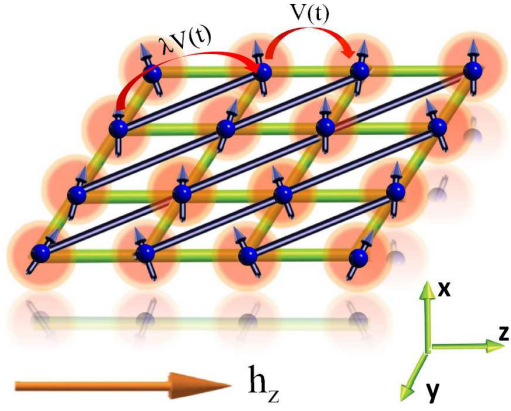}
\caption{(Color online)Sketch of a driven-dissipative magnetic system in square lattice with next-nearest neighboring coupling (blue bonds).} \label{fig:fig1}
\end{figure}

 \begin{figure}[htb]
\includegraphics[width=0.99\linewidth,bb=50 50 1000 549]{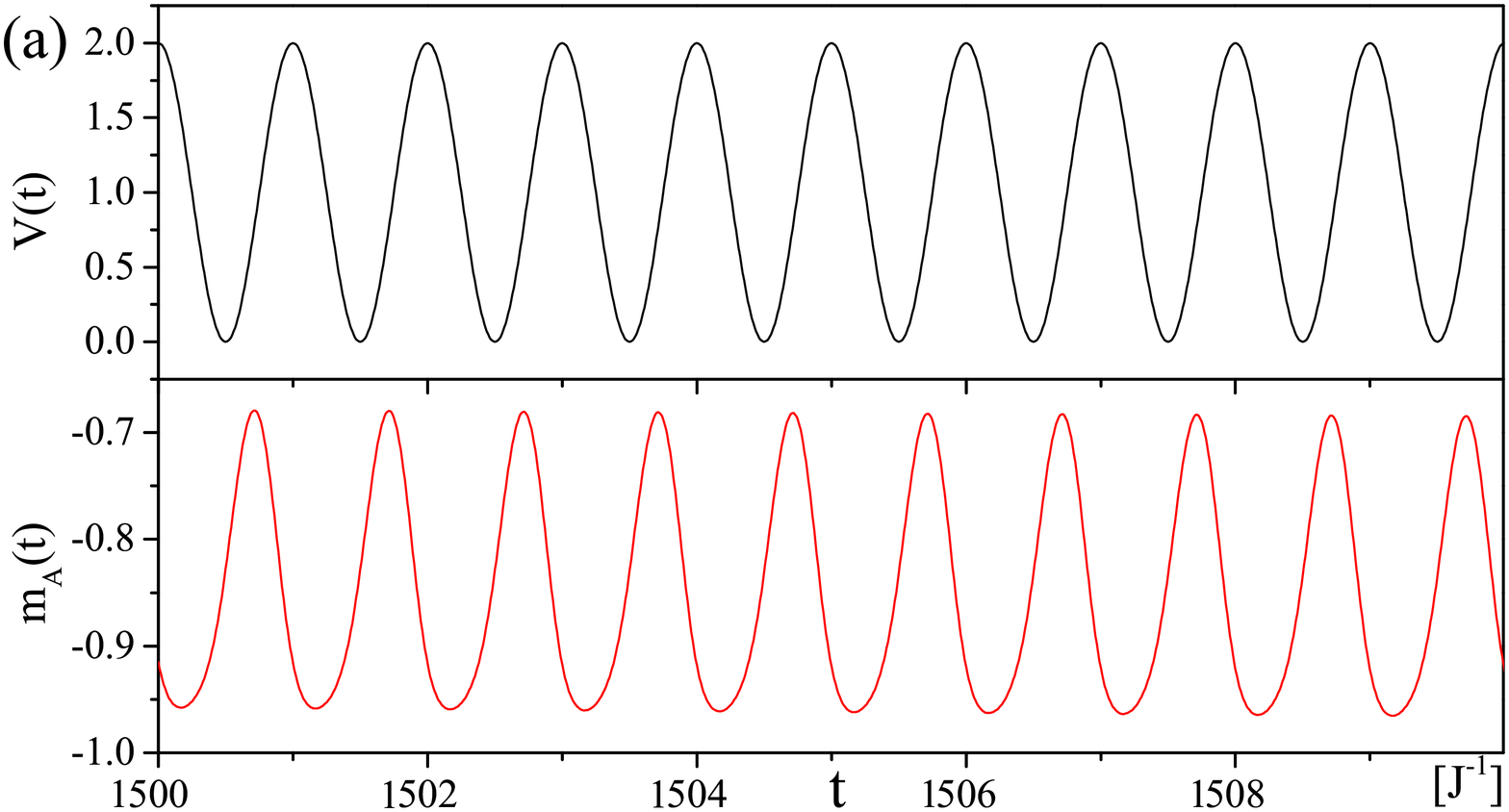}
\includegraphics[width=0.99\linewidth,bb=50 50 1000 549]{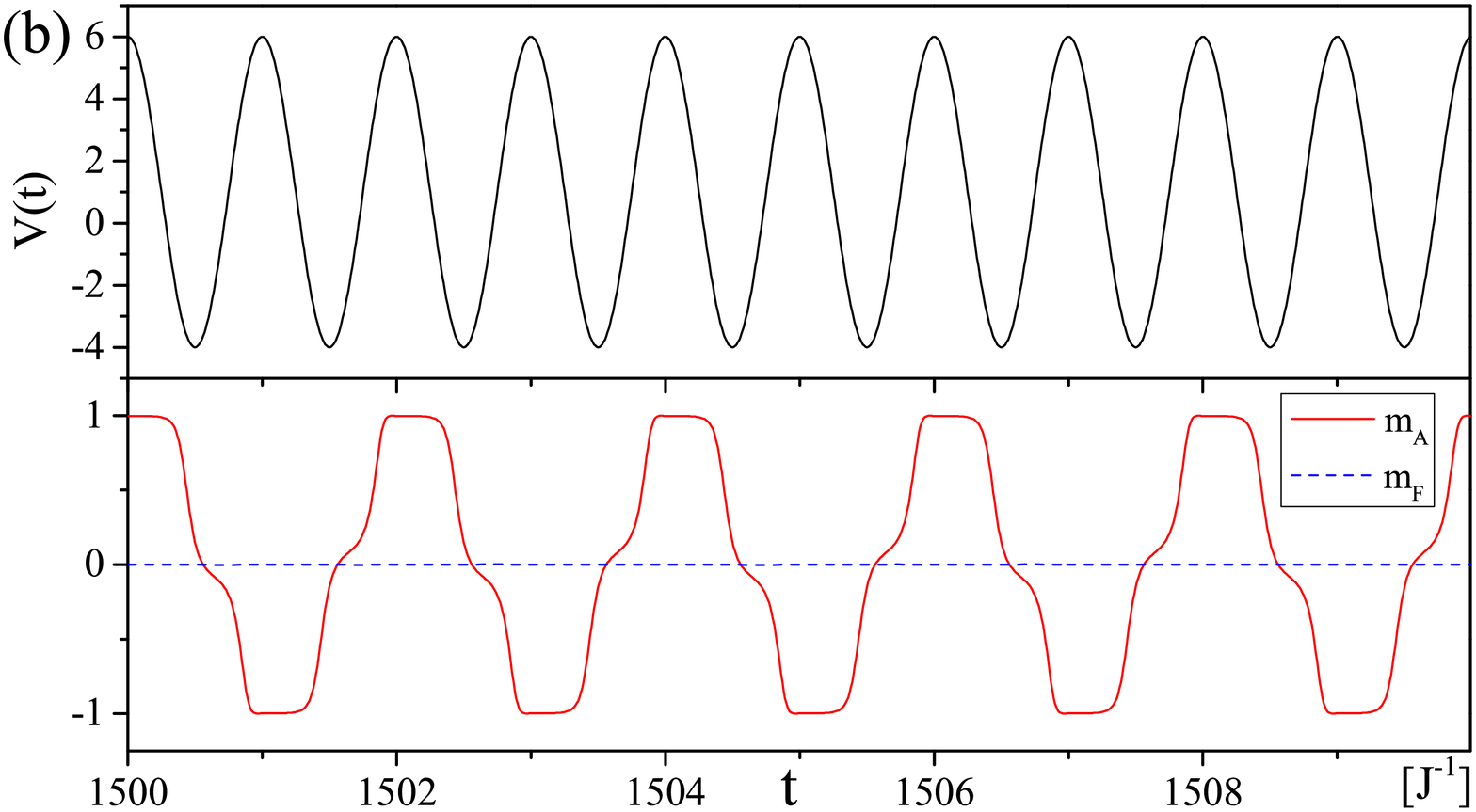}
\caption{(Color online)In the frustration-free case($\lambda=0$), (a)long-time dynamics of AF order parameter $m_A(t)$ with a weak driving ($J'=J$) and (b)long-time dynamics of the AF (red solid) and FM (blue dash) order parameters $m_A(t)$ and $m_F(t)$ with a strong driving ($J'=5J$). Parameters are chosen as $L=120$, $\eta=J$, $\mathcal{D}=0.01J$, $h_z=1.5J$ and $\omega=2\pi J$.} \label{fig:fig2}
\end{figure}

{\it Model and method --} We start with a two-dimensional classical spin model.  The system Hamiltonian reads:
\begin{equation}
H_s=H_0+\lambda H_f, \label{eq:Ham}
\end{equation}
where $H_0$ is a frustration-free Hamiltonian (transverse Ising model) defined in a $L\times L$ square lattice:
\begin{equation}
H_0(t)=V(t)\sum_{\langle ij\rangle} s_i^x s_j^x-\sum_i h_z s_i^z \label{eq:ham0}
\end{equation}
where the dynamical variable $\mathbf{s}_i$ is a three-dimensional classical vector with a fixed length $|\mathbf{s}_i|=1$, and the summation of $\langle ij\rangle$ is over the bonds  connecting two adjacent sites in the square lattice (green bonds, Fig.\ref{fig:fig1}). $h_z$ is the strength of the transverse field.  Periodic driving $V(t)=J+J'\cos\omega t$ is imposed on the interaction strength instead of the external field. $V(t)$ can be either positive or negative, corresponds to antiferromagnetic(AF) or ferromagnetic(FM) coupling. In our setup, alternate FM and AF couplings during the time evolution is crucial for our discussion. The frustration is introduced via the Hamiltonian $H_f$, whose strength is characterized by a dimensionless parameter $\lambda$. The frustration interaction is defined on one diagonal of each plaquette (blue bonds, Fig.\ref{fig:fig1}). The Hamiltonian $H_f$ reads:
\begin{equation}
H_f(t)=V(t)\sum_{\langle\langle ij\rangle\rangle} s_i^x s_j^x\label{eq:ham2}
\end{equation}
Only one diagonal of each plaquette is included because in the undriven case ($J'=0$) with $\lambda=1$, the Hamiltonian.(\ref{eq:Ham}) is reduced to an AF model defined on a triangle lattice,  a prototypical example of frustrated magnetism. Throughout this paper, we assume $J>0$ and fix the driving frequency as $\omega=2\pi$. In the absence of a thermal bath, the dynamics of each spin can be described by the equation of motion (EOM): $\dot{\mathbf{s}}_i=\mathbf{h}^0_i\times \mathbf{s}_i$, where the effective magnetic field $\mathbf{h}^0_i=[-V(t) \bar{s}_i^x, 0, h_z]$ with $\bar{s}_i^x= \sum_{\langle j\rangle} s_j^x+\lambda \sum_{\langle\langle j\rangle\rangle} s_j^x$.


The periodically-driven system is stabilized by introducing dissipation via coupling of each spin to a thermal bath, which can be modeled using methods familiar in the Brownian motion. The EOM for each spin is described by a stochastic Landau-Lifshitz-Gilbert equation\cite{Brown1963,Ma2010}:
\begin{equation}
\dot{\mathbf{s}}_i=\mathbf{h}_i\times \mathbf{s}_i- \eta  \mathbf{s}_i\times (\mathbf{s}_i\times \mathbf{h}_i) \label{eq:EOM}
\end{equation}
where $\eta$ is the dissipation strength  fixed as $\eta=J$ and  $\mathbf{h}_i=\mathbf{h}^0_i+\bm{\xi}_i(t)$ is the effective magnetic field, where $\bm{\xi}_i(t)$ is a three-dimensional zero-mean($\langle \xi_i^\alpha(t)\rangle_{\bm\xi}=0$) stochastic magnetic field  representing a thermal noise.  We further assume the local bath around different sites is independent of each other, and the stochastic variables satisfy: $\langle \xi_i^\alpha(t)\xi_j^\beta(t')\rangle_{\bm\xi}=\mathcal{D}^2\delta_{\alpha\beta}\delta_{ij} \delta(t-t')$
where $\alpha,\beta$ are the index of three space dimensions and $\mathcal{D}$ is the noise strength. The ensemble average $\langle\rangle_{\bm\xi}$ is over all the noise trajectories. For a thermal bath with temperature T, the strengths of the dissipation and noise satisfy the fluctuation-dissipation theorem
$\mathcal{D}^2=2T \eta$. We fix $\mathcal{D}=0.01J$, which corresponds to an extremely low temperature and does not plays a crucial role here. The high-temperature case is discussed in the supplementary material(SM)\cite{Supplementary}.

We discretize the stochastic differential Eq.(\ref{eq:EOM}) by adopting Stratonovich's formula, and solve it by the standard Heun  method\cite{Ament2016,Supplementary} with a time step of $\Delta t=10^{-3}J^{-1}$. In our simulation, we choose the initial state for each spin as $[s_i^x,0,s_i^z]$, where $s_i^x\in[-1,1]$ is a random number, and $s_i^z$ is fixed accordingly as $s_i^z=\sqrt{1-(s_i^x)^2}$ (the initial state dependence  has also been checked\cite{Supplementary}).  The system size in our simulation is up to $L=120$.  Despite the richness of the dynamical phase diagram of this driven-dissipative model, below, we will only focus on the dynamical phases with SSB in both space and time, and examine the roles of frustration and driving.

 \begin{figure}[htb]
\includegraphics[width=0.85\linewidth,bb=30 5 408 163]{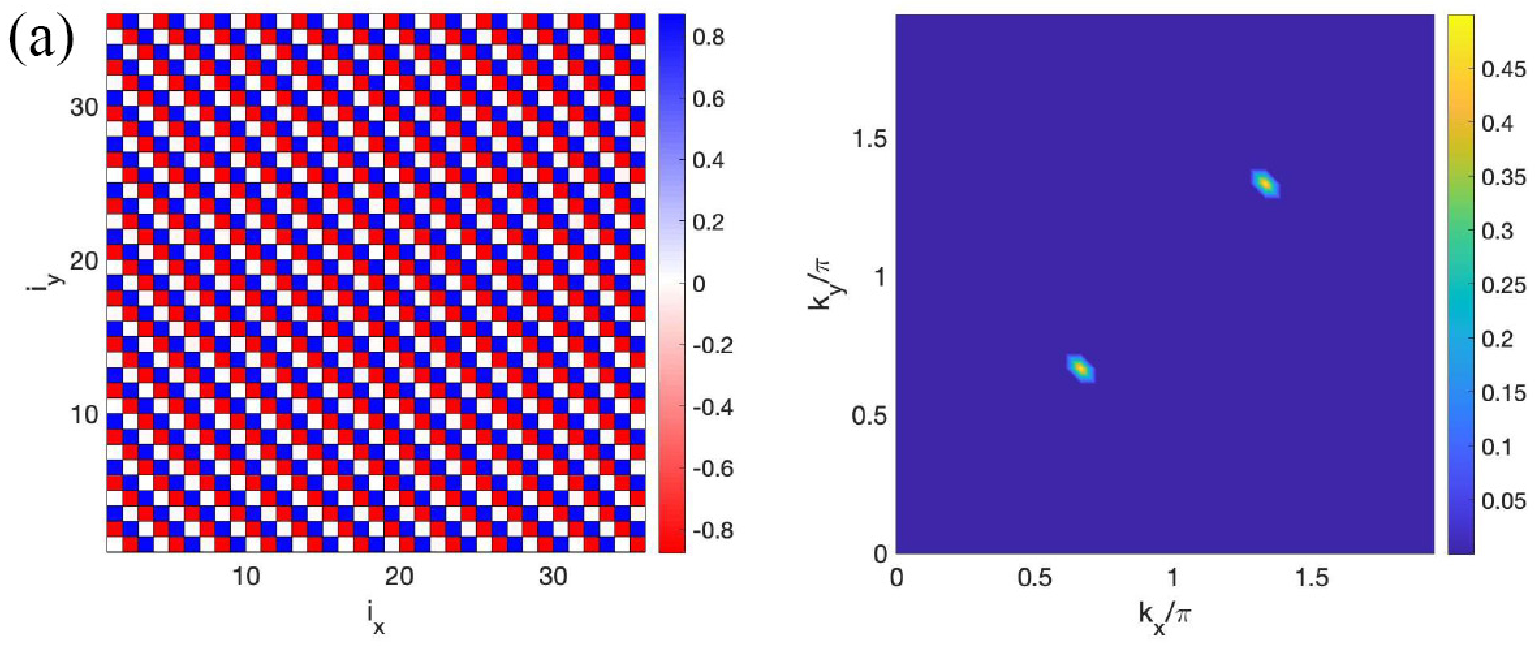}
\includegraphics[width=0.99\linewidth,bb=50 50 1000 549]{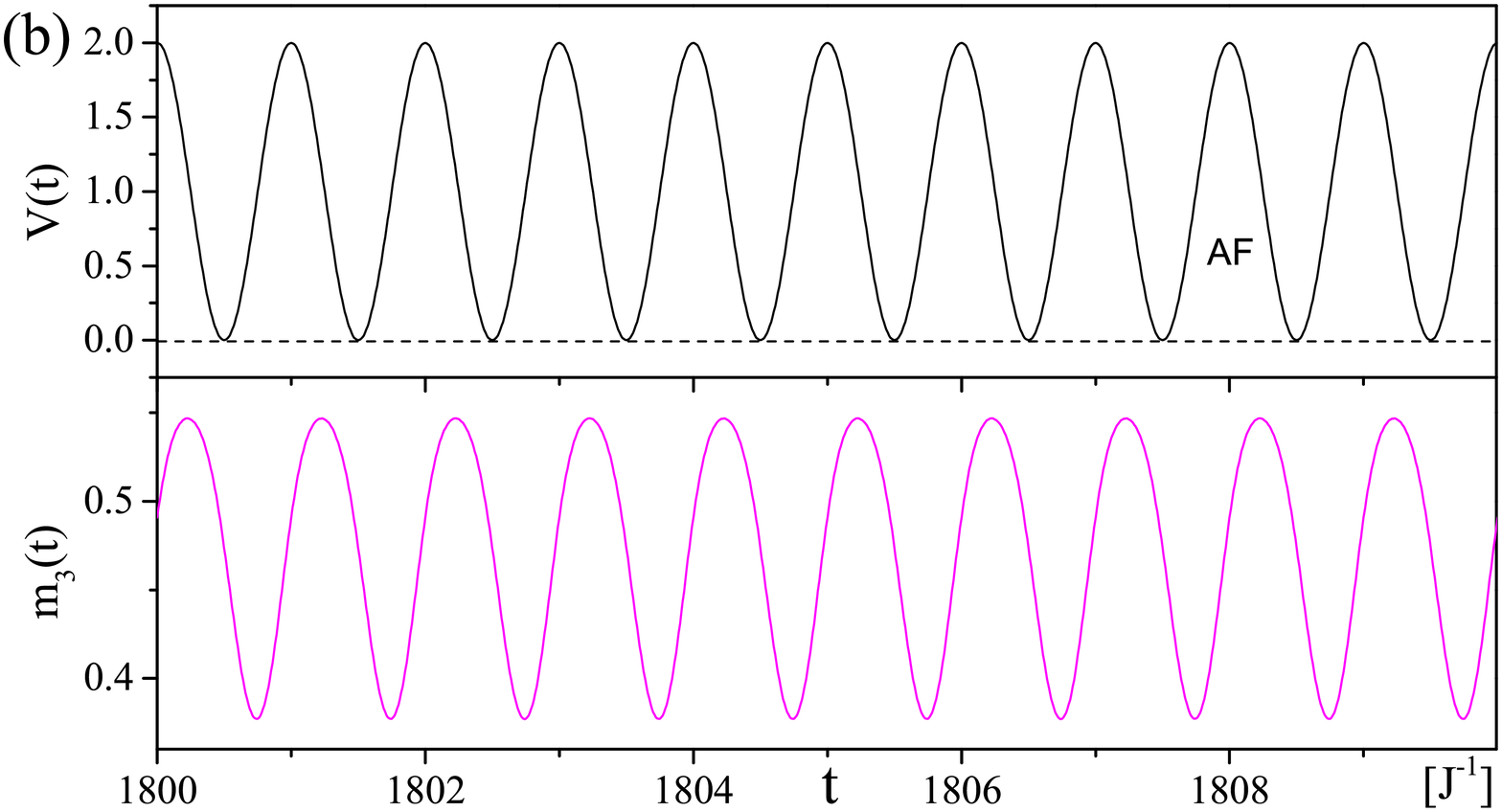}
\includegraphics[width=0.99\linewidth,bb=50 50 1000 549]{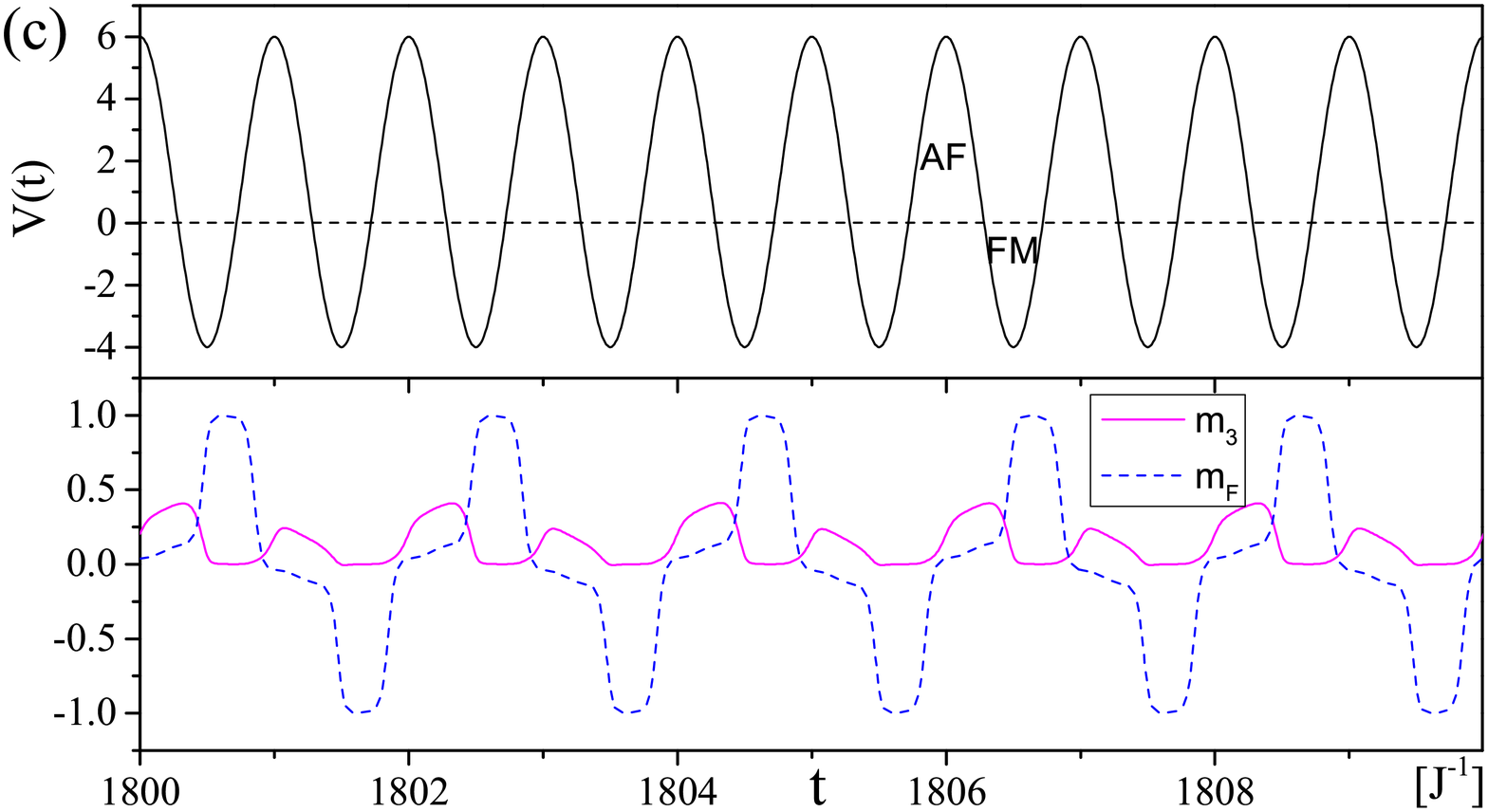}
\caption{(Color online)In a fully frustrated case ($\lambda=1$), (a) magnetization distribution $\{s_i^x\}$ of the long-time asymptotic state without driving ($J'=0$) in a real space (left panel) and its Fourier transformation with the peaks at momenta $\mathbf{Q}_0$ and $2\mathbf{Q}_0$ with  $\mathbf{Q}_0=(\frac {2\pi}3,\frac{2\pi}3)$ (right panel); (b)long-time dynamics of tripartite stripe order parameter $m_3(t)$ in the weakly driven case ($J'=J$); and (c)long-time dynamics of the tripartite stripe (purple solid) and the FM (blue dash) order parameters $m_3(t)$ and $m_F(t)$ in the strongly driven case ($J'=5J$).  $L=120$ except for (a) where $L=36$.  Other parameters are chosen the same as those in  Fig.\ref{fig:fig2}.} \label{fig:fig3}
\end{figure}

{\it Frustration-free case: an antiferromagnetic discrete time crystal (AF-DTC) --} We start with a frustration-free case ($\lambda=0$), where we monitor the magnetization dynamics  $s_\mathbf{i}^x(t)$  based on Eq.(\ref{eq:EOM}). Without driving($J'=0$), the system will relax to an equilibrium AF state with a nonzero order parameter $m_A=\frac 1{L^2}\sum_\mathbf{i} (-1)^{i_x+i_y} s_\mathbf{i}^x$. Fig.\ref{fig:fig2} indicates that such an AF order also persists in the presence of driving.  Fig.\ref{fig:fig2} (a) shows that in the weak driving case ($J'=J$) where the coupling is always AF ($V(t)\geq 0$), $m_A(t)$ finally oscillates around a finite value with a frequency synchronizing with driving.  At  strong driving ($J'=5J$), one can also observe a time-dependent AF order (Fig.\ref{fig:fig2} b), however, such an AF state fundamentally differs from its equilibrium counterpart in two aspects.  First, the long-range AF order is present at any time, even at those durations with FM coupling ($V(t)<0$). By contrast, the FM order parameter $m_F(t)=\frac 1{L^2}\sum_\mathbf{i}  s_\mathbf{i}^x(t)$ (the blue dash line, Fig.\ref{fig:fig2} b) vanishes during the whole evolution. Furthermore, different from the weakly driven case, $m_A(t)$ oscillates with a period doubling with respect to that of driving, thereby spontaneously breaking the discrete time translational symmetry from the symmetry group $\mathbb{Z}$ to $2\mathbb{Z}$. Consequently, such a state simultaneously breaks the space and time translational symmetry, thus it is an AF-DTC.

 \begin{figure*}[htb]
\includegraphics[width=0.33\linewidth,bb=28 51 650 550]{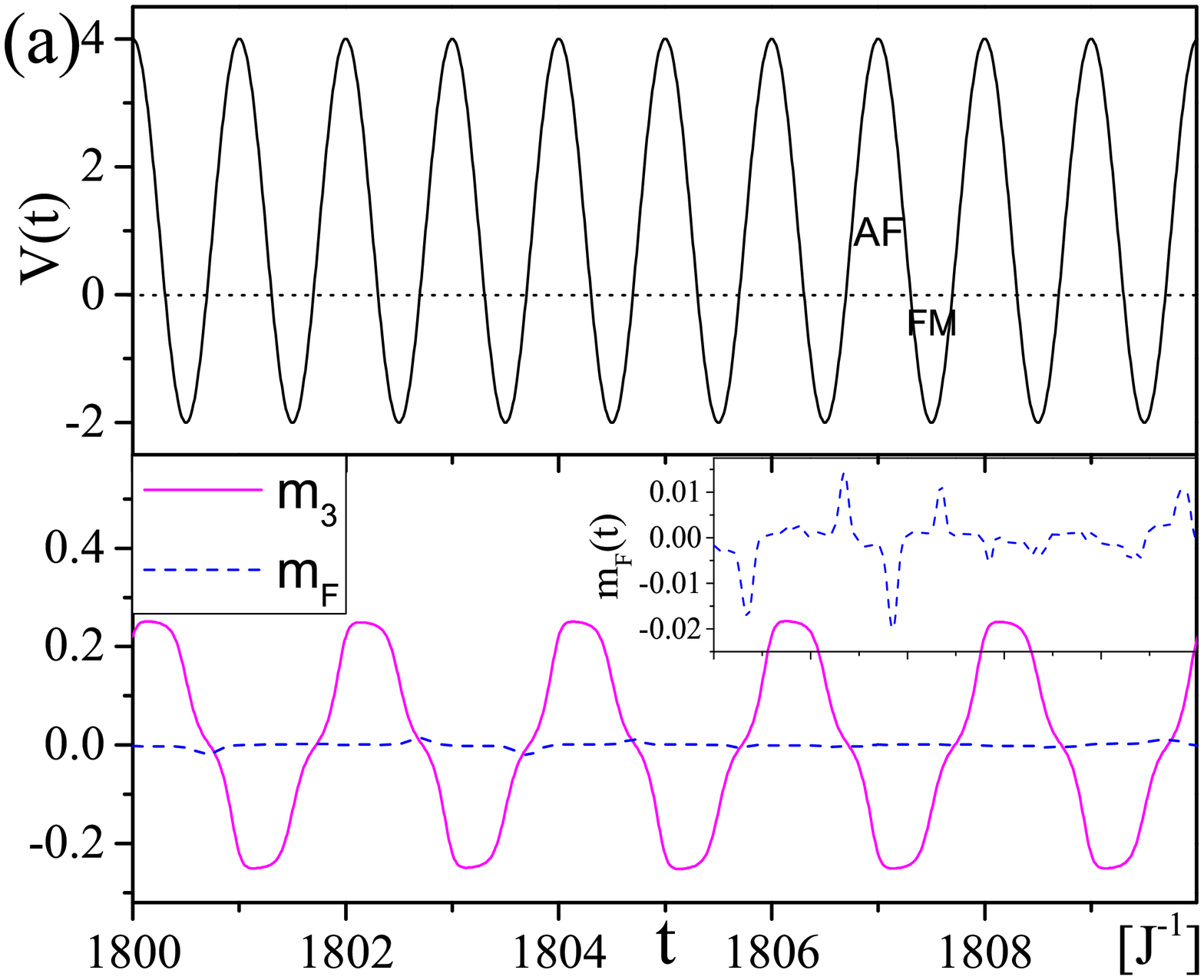}
\includegraphics[width=0.325\linewidth,bb=1 12 317 280]{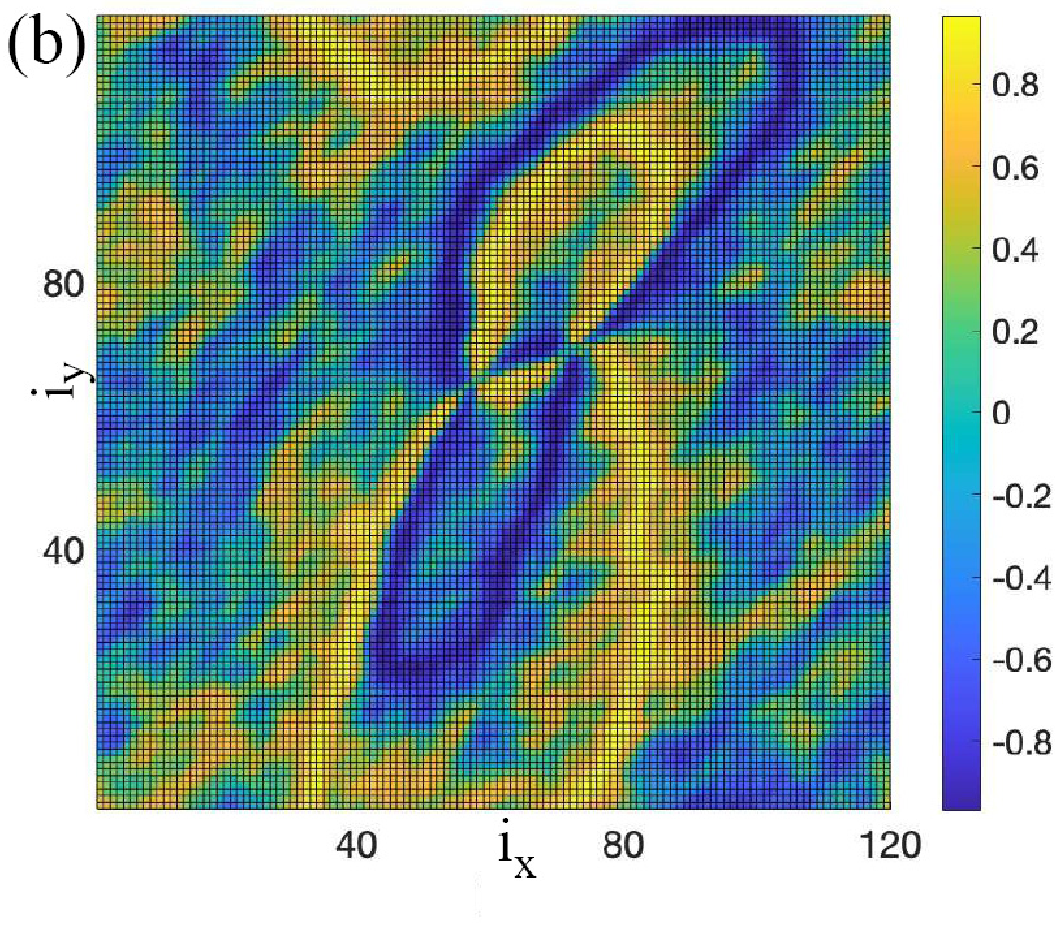}
\includegraphics[width=0.325\linewidth,bb=65 55 655 540]{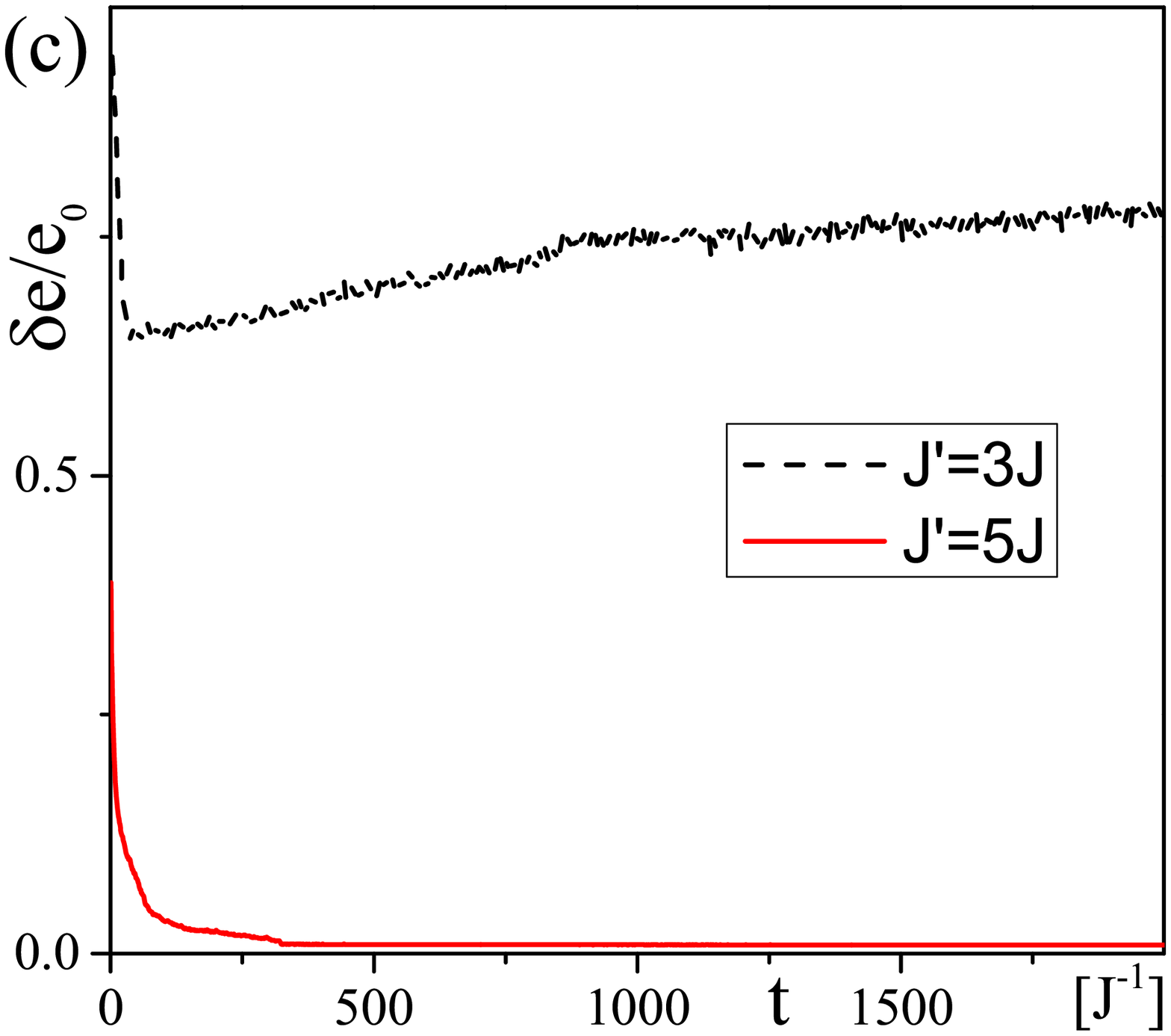}
\caption{(Color online)In a fully frustrated case ($\lambda=1$) with  intermediate driving ($J'=3J$),  (a) long-time dynamics of the tripartite stripe (purple solid) and FM (blue dash) order parameters $m_3(t)$ and $m_F(t)$ (inset magnifies $m_F(t)$), (b)typical magnetization distribution $\{s_i^x\}$ in a real space at an time slice with maximal FM coupling; and (c) comparison of the dynamics of excess energy with respect to the perfect FM state at the time slices with maximal FM coupling for the cases with intermediate ($J'=3J$) and strong  ($J'=5J$) driving.  Other parameters are chosen the same as in Fig.\ref{fig:fig2}. } \label{fig:fig4}
\end{figure*}

{\it Fully frustrated case: A zoo of nonequilibrium phases of matter --} Now we consider a fully frustrated  case with $\lambda=1$, where the lattice is equivalent to a triangle lattice (a study of intermediate frustration with $0<\lambda<1$ can be found in SM\cite{Supplementary}). In equilibrium magnetism, frustration works against AF order. One may wonder whether it plays a similar role of suppressing the forementioned AF-DTC order in this non-equilibrium setup.  If so, which order does frustration suppress? Is it the spatial AF or temporal DTC, or both? What kinds of space-time structures will emerge once the AF-DTC is destroyed? In the absence of driving ($J'=0$),  the system  will relax toward an equilibrium state close to the ground state of the Hamiltonian.(\ref{eq:Ham}). The transverse field distinguishes our model from the pure Ising model in the triangle lattice with extensive ground state degeneracy. Fig.\ref{fig:fig3} (a) depicts the steady state magnetization with a tripartite structure and a stripe order along the diagonal direction. This magnetic order is characterized by the order parameter $m_3(t)=\frac 1{L^2}\sum_\mathbf{i} \sin[\mathbf{Q}_0\cdot\mathbf{i}]s_\mathbf{i}^x(t)$ with $\mathbf{Q}_0=(\frac {2\pi}3,\frac{2\pi}3)$.

The stripe order parameter $m_3(t)$ starts to oscillate once the periodic driving is switched on.  With weak driving ($J'=J$), Fig.\ref{fig:fig3} (b) shows that $m_3(t)$ oscillates around a finite value with a period the same as driving, thus the tripartite stripe order still persists in this nonequilibrium case. For strong driving ($J'=5J$), $V(t)$ alternates between FM and AF coupling during the evolution, thereby significantly changing the dynamics. Fig.\ref{fig:fig3} (c) illustrates that at the stripe/FM long-range order is built during AM/FM coupling.  However, this does not mean that the system adiabatically follows the instantaneous ground state of Hamiltonian.(\ref{eq:Ham}). Instead, it is in a genuine nonequilibrium state because both $m_3(t)$ and $m_F(t)$ develop a DTC order in the time domain, indicating that a temporal correlation is dynamically built. In other words, the asymptotic state with strong driving is a space-time crystal that simultaneously breaks not only the space and time translational symmetry, but also the $Z_2$ symmetry of the Hamiltonian.(\ref{eq:Ham}) in spin space. This space-time crystal is significantly different from the AF-DTC phase observed in the frustration-free case, where the FM order is absent even during FM coupling.

The situation is more interesting with intermediate coupling ({\it e.g.} $J'=3J$). Fig.\ref{fig:fig4} (a) illustrates that the tripartite stripe and associated DTC orders still persist. However, different from that in the strongly driven case, the long-range FM order is not built during the whole period. Its order parameter $m_F(t)$ (dash blue line,Fig.\ref{fig:fig4} a) stochastically oscillates with a small amplitude that decreases with system size\cite{Supplementary}. A typical magnetization configuration  $\{s_i^x\}$ at a time slice with a maximum FM coupling is plotted in Fig.\ref{fig:fig4}(b), which exhibits a wealth of FM domain walls(DW). To measure the density of the DWs, we define an excess energy density with respect to the perfect FM state as $\delta e(t)=\frac 1{L^2}\langle H^I_s(t)\rangle-e_0$, where $\langle H^I_s(t)\rangle$ is the instantaneous interacting energy at time t (we only focus on the time slices with maximal FM coupling) and $e_0$ is the energy density of a perfect FM state along x-direction. In Fig.\ref{fig:fig4} (c), for $J'=3J$, $\delta e(t)/e_0$ ultimately saturates toward a large value, indicating that it is far from a perfect FM state and the density of DWs does not decay in time.  By contrast, for the strong driving case with $J'=5J$, excess energy  is very small, indicating that the system could reach an almost perfect FM state at the maximal FM coupling. The results show that at intermediate coupling, even though the system builds a short-range FM correlation, it has no time to develop long-range FM order before the coupling turns back to AF within a driving circle. The difference between the intermediate and strong coupling cases is because the FM coupling duration in the former is shorter than that in the latter. The FM duration is not long enough for the system to build up long-range FM correlation.

{\it Discussion --} We present  several tunable parameters in our model; hence, one may wonder whether the forementioned space-time crystals require fine-tuned parameters in the massive parameter space. We will answer this question by discussing the role of different model parameters. As for the bath, the dissipation parameter $\eta$ is introduced to stabilize the system, thereby controlling the transient relaxation dynamics.  Small thermal noises ($\mathcal{D}\ll J$) are introduced to avoid the system from being trapped in the metastable states. It will not change the nature of the space-time crystals (Note, however, that a strong thermal fluctuation could melt the space-time crystals\cite{Supplementary,Yue2022}). Including a transverse field ($h_z$) makes the spin dynamics nontrivial, otherwise it is only a precession along the x-direction.

The driving amplitude $J'$, frequency $\omega$ and the frustration strength $\gamma$ are crucial in determining the different space-time patterns in long-time asymptotic states. As previously discussed, the DTC order can only emerge when the coupling $V(t)$ alternates between  AF and FM during the evolution. With this, a sufficiently large $J'$ is required.  As for $\omega$, in the limit $\omega\gg J$, a fast periodic driving only manifests itself via its mean value thus doesn't play a key role here. In the opposite limit where the driving period is much longer than the relaxation time $\omega\ll  \eta$, the system is always close to the thermodynamical equilibrium states with FM or stripe orders depending on the sign of $V(t)$. However, these FM/stripe order parameters cannot spontaneously organize themselves into a DTC in the time domain because in such an adiabatic limit, the time slices with maximal AF and FM coupling are separated so far that a temporal correlation cannot be built up between them\cite{Supplementary}.  Therefore, to realize nontrivial space-time patterns,  periodic driving must be slow enough for the system to react accordingly and develop different magnetic orders in space, but not too slow so these magnetic order parameters could build a long-range (DTC) correlation in time domain. In short, the realization of the space-time crystals requires a strong driving amplitude and an intermediate driving frequency. It is a robust phase in the phase diagram.


The frustration in our model suppressed the AF order and induced other magnetic states, similar as it does in  equilibrium magnetism. However, in our nonequilibrium setup where $V(t)$ keeps changing its sign, frustration may facilitate an FM order which is absent in the frustration-free case. This is because in during FM coupling ($V(t)<0$), the Hamiltonian.(\ref{eq:ham2}) no longer leads to ``frustration'', instead it increases the effective FM coupling, while during AF coupling, frustration still suppress the AF order. As a consequence, frustration makes the AF order observed in the frustration-free case give way to a FM order. This  can be seen more clearly in an intermediate frustration regime ($\lambda\in[0.68,0.9]$), where we find a DTC phase with only FM  but no stripe order\cite{Supplementary}.

{\it Conclusion and outlook --} In this work, we introduce frustration into a driven-dissipative magnetic system, and show that it gives rise  to a wealth of nonequilibrium phases  with intriguing SSB in both space and time.

 Future developments will include the generalization of  these results into models with different lattice and spin symmetries.  For instance, in other geometric frustrated lattices ({\it e.g.} kagome or pyrochlore), one may expect nonequilibrium phases with other magnetic patterns ({\it e.g.} nematic or spin ice) in space and temporal orders (e.g. algebraic temporal correlation\cite{Svistunov2018}) in time. A more exciting possibility is the realization of  nonequilibrium states with intertwined space-time symmetries that cannot be decomposed into a direct product of spatial and temporal symmetries\cite{Xu2018,Gao2021}. As for the spin symmetry, the Hamiltonian.(\ref{eq:Ham}) preserves the Ising symmetry, generalizing the spin symmetry to continuous ones ({\it e.g.} $U(1)$) may lead to intriguing non-equilibrium phenomena ({\it e.g.}  a Berezinskii-Kosterlitz-Thouless-like phase transition in such a driven-dissipative system, where the traditional binding-unbinding picture of vortex\cite{Kosterlitz1973} may be modified in the context of nonequilibrium physics\cite{Altman2015}.

 A more fundamental question here is  to understand the ``excitations'' in such space-time crystals,  {\it e.g.} its response to external stimulation.  In equilibrium physics, elementary excitation is closely related with SSB ({\it e.g.} Goldstone theorem), one may wonder whether similar relation holds for these nonequilibrium phases with intriguing SSB\cite{Yang2021}. If so, how these excitations ({\it e.g.} nonequilibrium counterparts of magnon, soliton or phonon) are characterized and realized?  This problem is not only of fundamental interest, but also of immense practical importance considering its relevance to the experimental observable effect of these nonequilibrium phases.

{\it Acknowledgments}.---This work is supported by the National Key Research and Development Program of China (Grant No. 2020YFA0309000), NSFC of  China (Grant No.12174251), Natural Science Foundation of Shanghai (Grant No.22ZR142830),  Shanghai Municipal Science and Technology Major Project (Grant No.2019SHZDZX01). ZC thank the sponsorship from Yangyang Development Fund.


\newpage

\begin{center}
\textbf{\Large{Supplementary material}}
\end{center}

In this supplementary material, we first provide some details of the Heun algorithm as well as a benchmark of the numerical results.  Then we numerically check the convergence of our results with the finite discrete time step $\Delta t$ and the dependence of our results on the system size, noise trajectories, and initial states.  The role of the driving frequency, frustration and thermal fluctuation   is also discussed.
\section{Details about the numerical simulation}

\subsection{Heun algorithm}
Here, we first derive the discrete formalism of  stochastic Landau-Lifshitz-Gilbert (LLG) equation based on Stratonovich's formula, then formulate the Heun algorithm to solve this  Stochastic Differential equations (SDE). A stochastic LLG equation reads:
\begin{equation}
\dot{\mathbf{s}}_i=\frac 1{\eta^2+1}[\mathbf{h}_i\times \mathbf{s}_i- \eta  \mathbf{s}_i\times (\mathbf{s}_i\times \mathbf{h}_i)] \label{eq:EOM}
\end{equation}
where $\mathbf{s}_i$ is a unit vector.  $\mathbf{h}_i(t)=\mathbf{h}^0_i(t)+\bm{h}^T_i(t)$ is the effective magnetic field. $\mathbf{h}^0_i(t)$ comes from the interaction between spin i and its neighbors.  $\bm{h}^T_i(t)$ is a three-dimensional random magnetic field representing the thermal noise, which satisfies:
\begin{eqnarray}
\langle h_i^{T\alpha}(t)\rangle_{\bm\xi}&=&0\\
\langle h_i^{T\alpha}(t)h_j^{T\beta}(t')\rangle_{\bm\xi}&=&\mathcal{D}^2\delta_{\alpha\beta}\delta_{ij} \delta(t-t') \label{eq:noise}
\end{eqnarray}
where $\alpha,\beta$ are the index of three spatial dimensions  and $\mathcal{D}$ is the strength of the noise. $\langle\rangle_{\bm\xi}$ is the ensemble average over all the trajectories of noises. According to the Fluctuation-dissipation theorem, the strength of the thermal noise and the dissipation satisfies  the relation:
\begin{equation}
\mathcal{D}^2=2T \eta. \label{eq:FDT}
\end{equation}

To solve this SDE numerically, we first discretize the time with the time step of $\Delta t$.  Let the spin configuration in the $m$-th time step ($t_m=m\Delta t$) be $\{\mathbf{s}_i^{m}\}$,  the calculation of   $\{\mathbf{s}_i^{m+1}\}$ can be divided into two steps  in the Heun algorithm.

In the first step,  we derive an intermediate spin configuration $\{\tilde{\mathbf{s}}_i^{m+1}\}$:
\begin{equation}
\tilde{\mathbf{s}}_i^{m+1}=\mathbf{s}_i^{m}+\frac 1{\eta^2+1}[\mathbf{h}^{m}_i\times \mathbf{s}_i^m- \lambda  \mathbf{s}_i^{m}\times (\mathbf{s}_i^{m}\times \mathbf{h}^{m}_i)]\Delta t \label{eq:mid}
\end{equation}
with $\mathbf{h}_i^m=\mathbf{h}^0_{i,m}+\bm{\tilde{h}}^T_{i,m}$, where $\mathbf{h}^0_{i,m}=\mathbf{h}^0_i(t_m)$ and $\bm{\tilde{h}}^T_{i,m}$ is a stochastic magnetic field satisfying:
\begin{equation}
\tilde{h}^{T\alpha}_{i,m}=\frac{\mathcal{D}}{\sqrt{\Delta t}}\xi_{i,m}^\alpha \label{eq:random}
\end{equation}
where $\xi_i^\alpha$ is a random number satisfying the Gaussian distribution with zero mean and unit variance: $\langle \xi_i^\alpha\rangle_{\bm\xi}=0$, $\langle (\xi_i^\alpha)^2\rangle_{\bm\xi}=1$.

In the Heun algorithm, $\mathbf{s}_i$ at the $m+1$-th time step can be expressed as:
\begin{eqnarray}
\nonumber &\mathbf{s}_i^{m+1}&=\mathbf{s}_i^{m}+\frac{\Delta t}2\{ \mathbf{h}^{m}_i\times \mathbf{s}_i^m- \lambda  \mathbf{s}_i^{m}\times (\mathbf{s}_i^{m}\times \mathbf{h}^{m}_i)\\
&+&\tilde{\mathbf{h}}^{m+1}_i\times \tilde{\mathbf{s}}_i^{m+1}- \lambda  \tilde{\mathbf{s}}_i^{m+1}\times (\tilde{\mathbf{s}}_i^{m+1}\times \tilde{\mathbf{h}}^{m+1}_i)\}
\end{eqnarray}
where $\tilde{\mathbf{s}}_i^{m+1}$ has been defined in Eq.(\ref{eq:mid}), and $\tilde{\mathbf{h}}_i^{m+1}=\mathbf{h}^0_{i,m+1}+\bm{\tilde{h}}^T_{i,m}$.

 \begin{figure*}[htb]
\includegraphics[width=0.325\linewidth,bb=80 50 770 545]{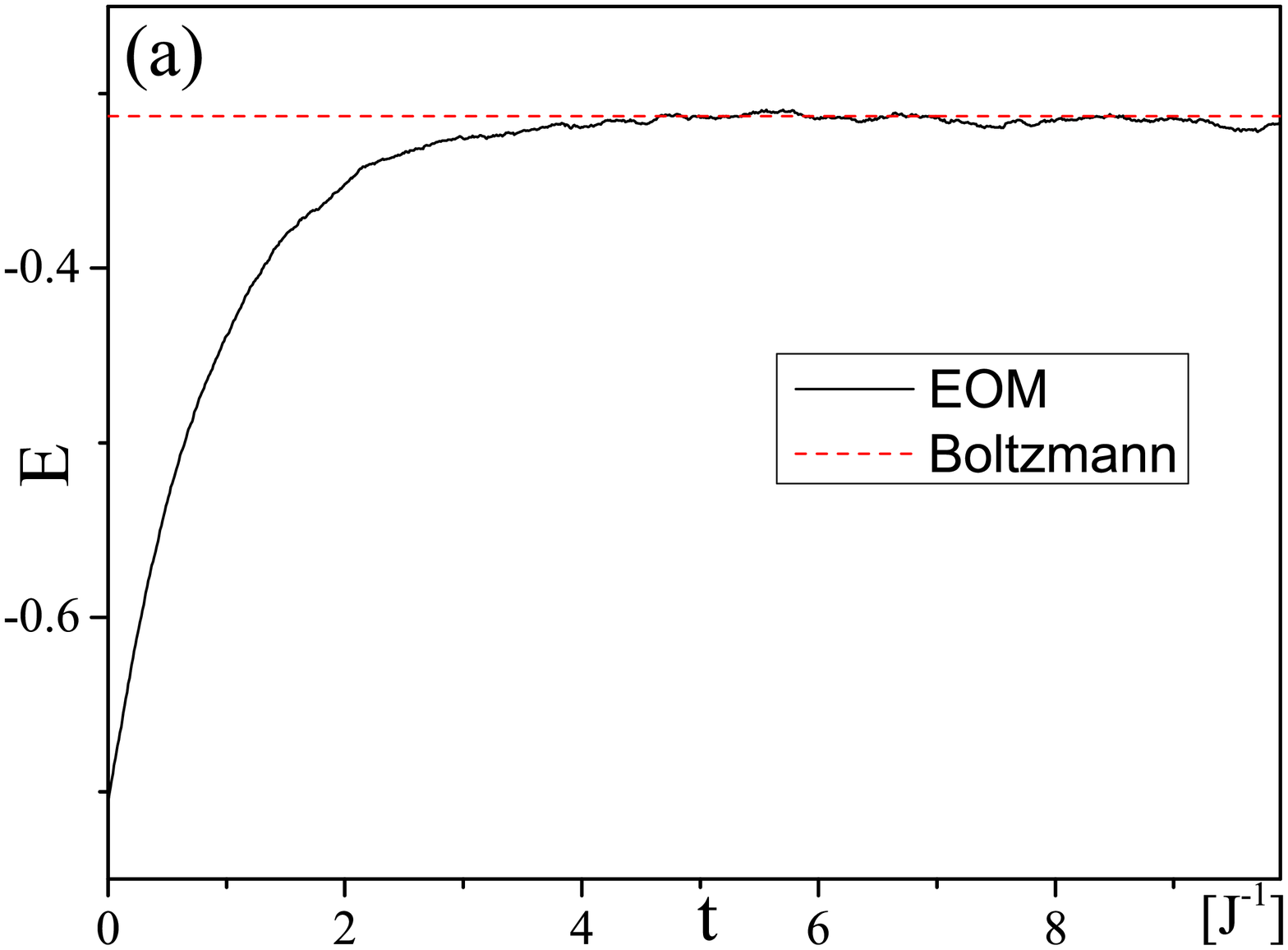}
\includegraphics[width=0.325\linewidth,bb=80 50 770 545]{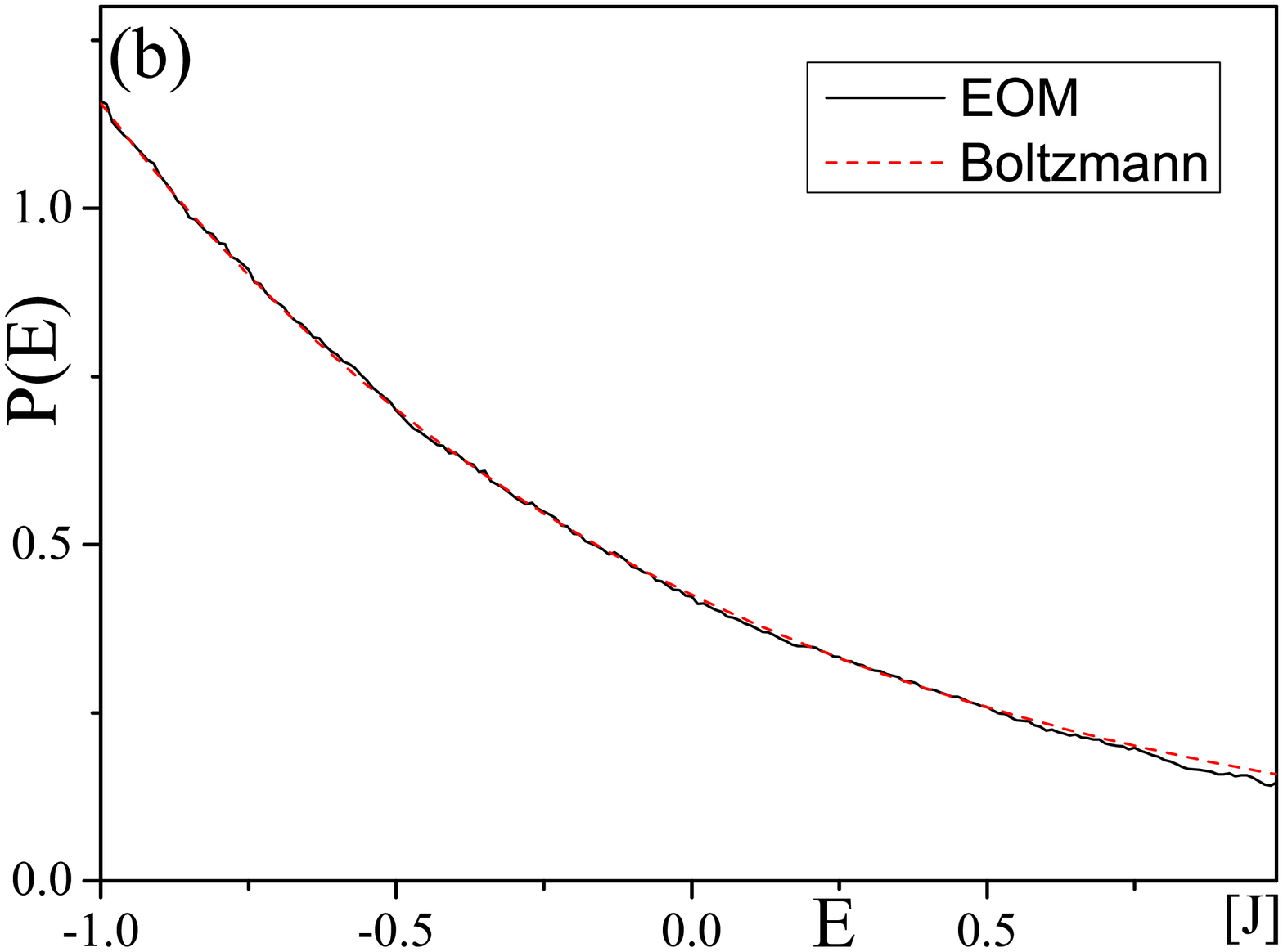}
\includegraphics[width=0.325\linewidth,bb=80 50 770 545]{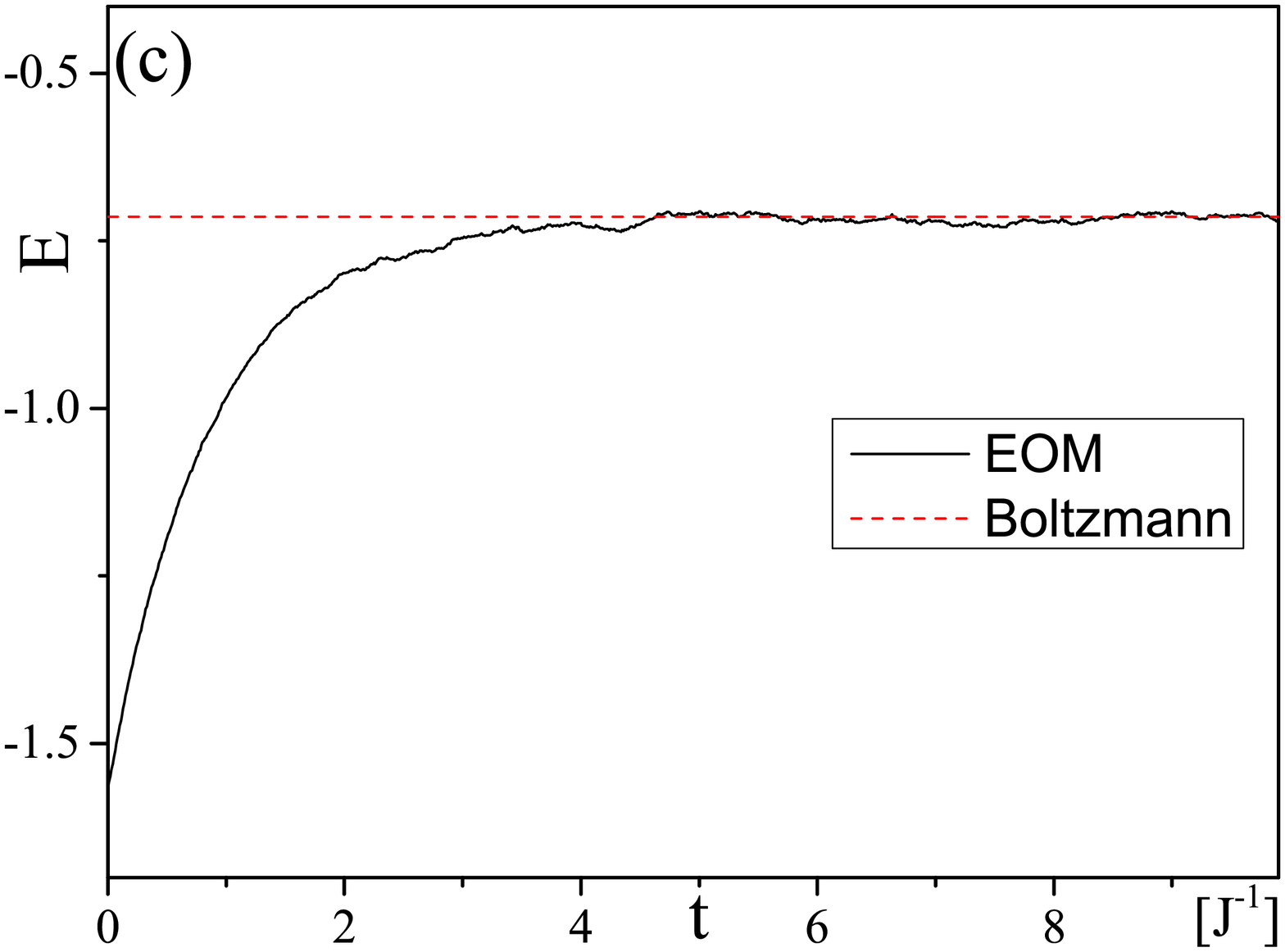}
\caption{(Color online) Relaxation dynamics  predicted by  Eq.(\ref{eq:EOM})  in the absence of driving. (a) the dynamics of the energy for a single spin model predicted by the EOM.(\ref{eq:EOM}), whose long-time asymptotic value approaches that predicted by the statistical ensemble average; (b) the distribution of energy predicted by the EOM.(\ref{eq:EOM}), which agrees very well with the Boltzmann distribution. (c) The dynamics of the energy for a two-spin model predicted by the EOM.(\ref{eq:EOM}) and the value predicted by statistical ensemble average (red dashed line). The parameters are chosen as $h_x=J$ , $\eta=J$ and the temperature $T=J$.} \label{fig:SM1}
\end{figure*}

 \begin{figure*}[htb]
\includegraphics[width=0.325\linewidth,bb=80 50 770 545]{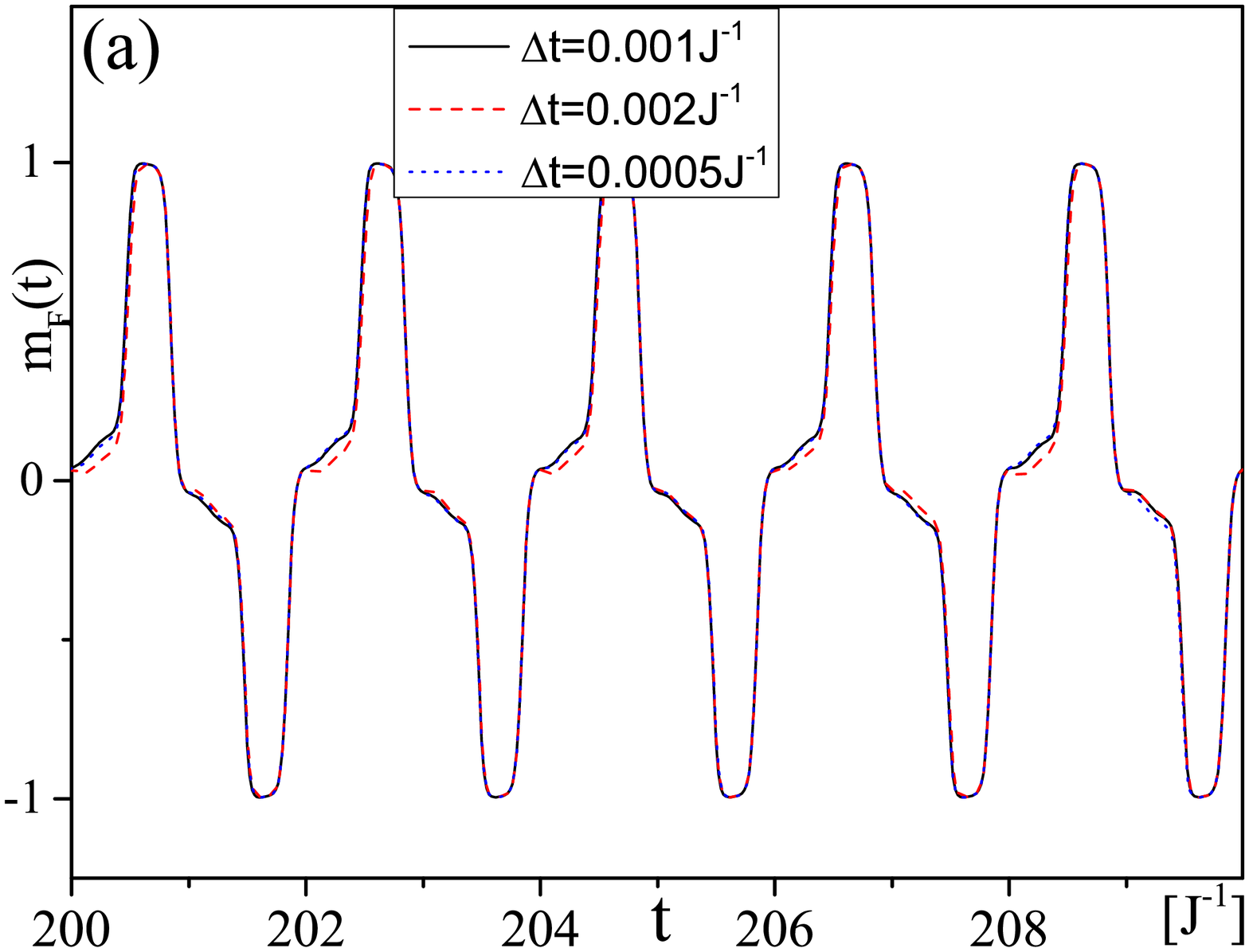}
\includegraphics[width=0.325\linewidth,bb=80 50 770 545]{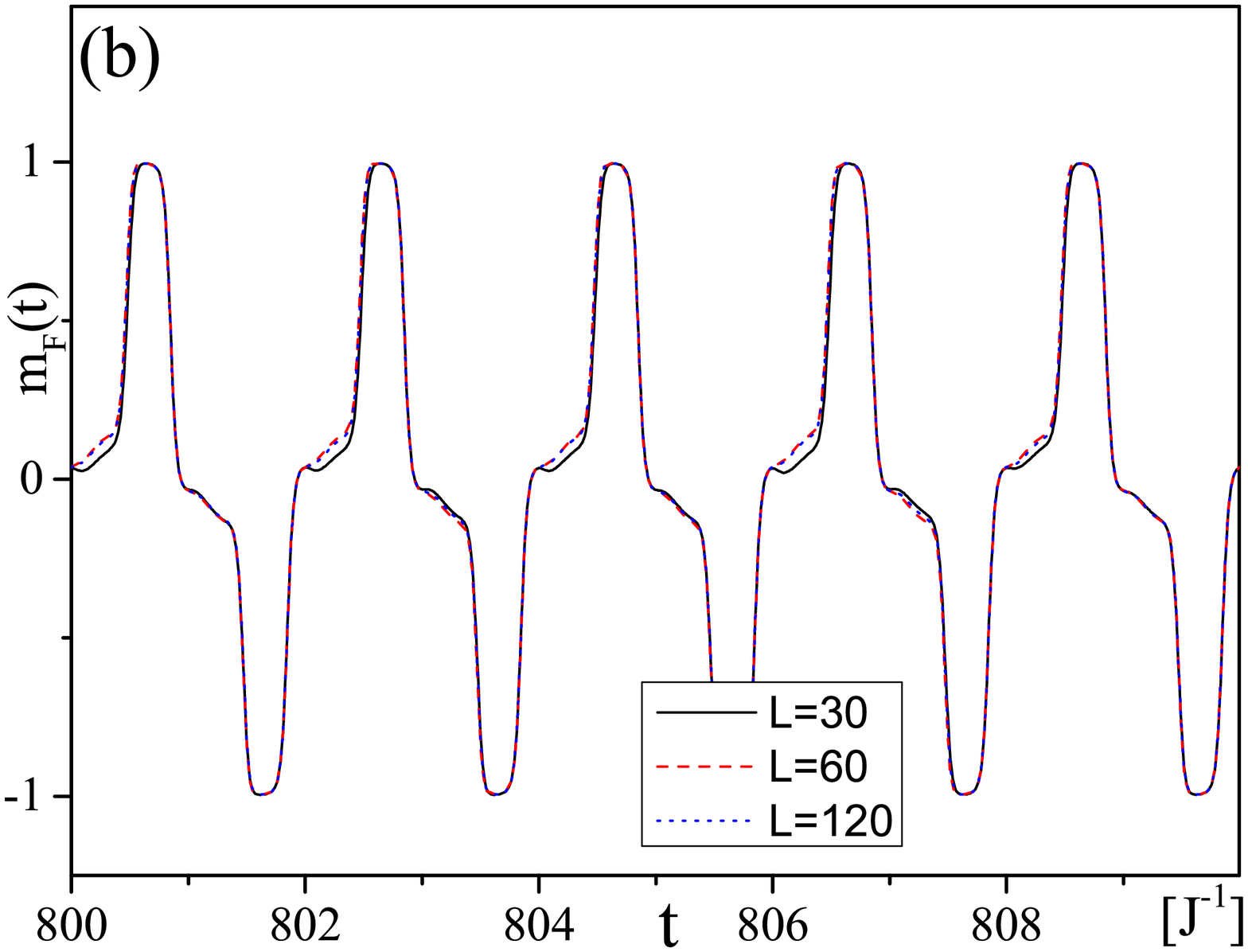}
\includegraphics[width=0.325\linewidth,bb=80 50 770 545]{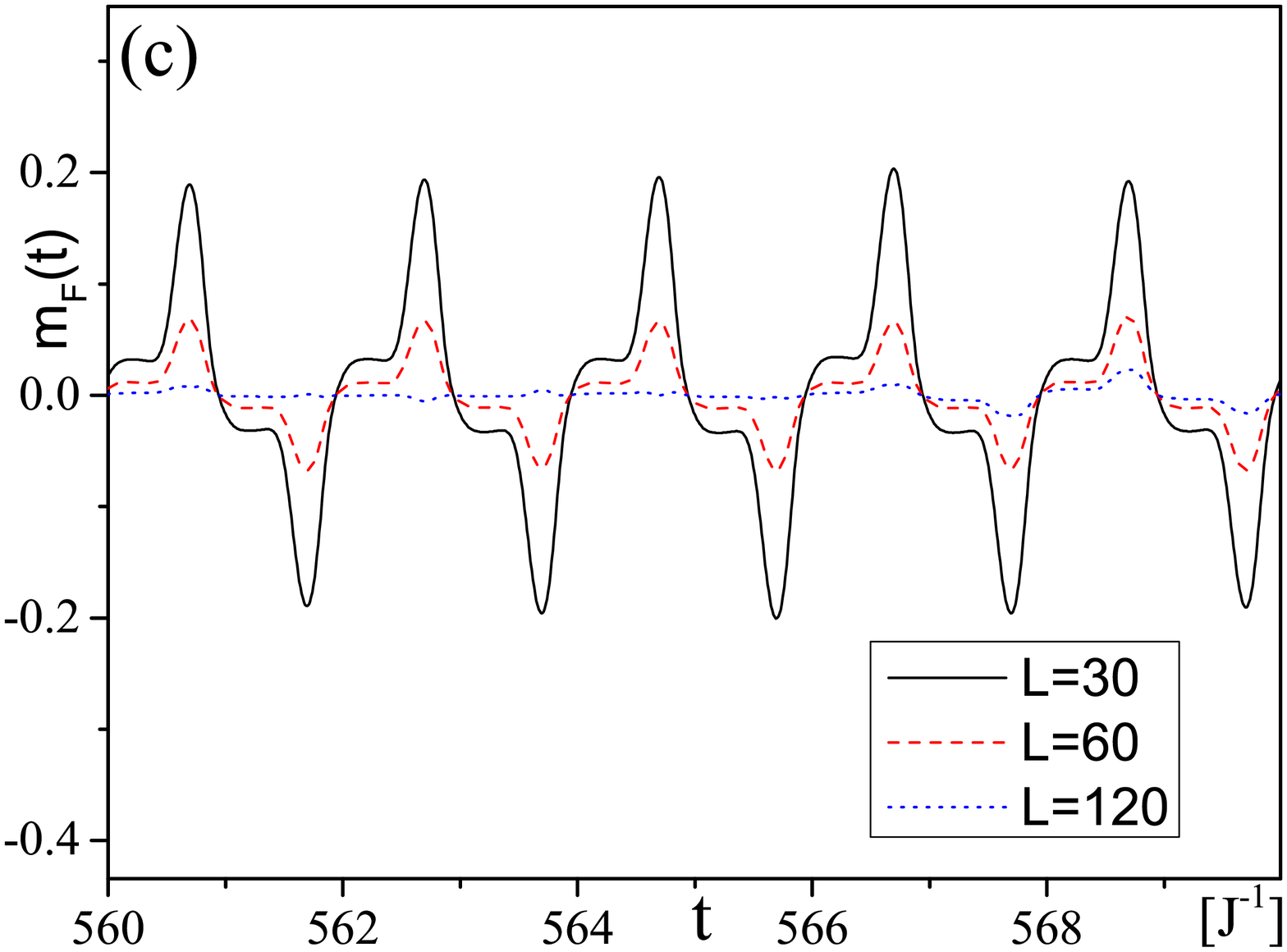}
\caption{(Color online) (a) Comparison between the dynamics of FM order parameter $m_F(t)$ with different $\Delta t$ with $J'=5J$, $L=30$. Comparison between the dynamics of FM order parameter $m_F(t)$ with different system size $L$ (b) in the strong driving case ($J'=5J$) and (c) in the intermediate driving case ($J'=3J$). Other parameters are chosen as $\eta=J$, $\mathcal{D}=0.01J$,$\lambda=J$, $h_z=1.5J$,$\omega=2\pi$. } \label{fig:SM2}
\end{figure*}

\subsection{Benchmark: spin models without driving}
It is known that once a system couples to a heat bath, it will finally relax to a thermodynamical equilibrium state at the same temperature of the bath, irrespective of its initial state. To verify the validity of the EOM.(\ref{eq:EOM}), we consider two simple spin models as benchmarks, which show that the long-time asymptotic state derived by EOM.(\ref{eq:EOM}) is indeed the thermodynamic equilibrium states at a temperature determined by  Eq.(\ref{eq:FDT}).

The first model we consider is a single spin model with the Hamiltonian:
\begin{equation}
H^s_1= h_z s^z
\end{equation}
By solving EOM.(\ref{eq:EOM}) using the Heun algorithm, we can find the energy of the system $E(t)=\langle H(t)\rangle$ quickly relaxes to a value with small statistical fluctuation.   According to the statistical physics, for a thermodynamical equilibrium state, the long-time average of the system energy is supposed to be the same with the value predicted by the statistical ensemble, which is
\begin{equation}
E_s=\frac 1Z\int_0^\pi \sin\theta d\theta [h_z \cos\theta] e^{-\beta h_z\cos\theta}
\end{equation}
where $\theta$ is the angle between the spin vector and z-axis, and $Z=\int_0^\pi \sin\theta d\theta  e^{-\beta h_z\cos\theta}$ is the partition function. As shown in Fig.\ref{fig:SM1} (a), the time average value of $\langle H(t)\rangle$ agrees very well with the ensemble averaged value $E_s$ within the statistical error. In addition, one can study the statistical distribution of $E(t)$ during the long-time dynamics, $P(E)$ agrees very well with the Boltzmann distribution, as shown in Fig.\ref{fig:SM1} (b).

We also check a two-spin model with the Hamiltonian:
\begin{equation}
H_s^2=J s_1^x s_2^x+h_z(s_1^z+s_2^z) \label{eq:spin2}
\end{equation}
We study the evolution of the system, and focus on its energy.  As shown in Fig.\ref{fig:SM1} (c), in the long-time limit, the system energy will approach the value predicted by canonical ensemble accompanied by small statistical fluctuations.

 \begin{figure}[htb]
\includegraphics[width=0.9\linewidth,bb=80 50 770 545]{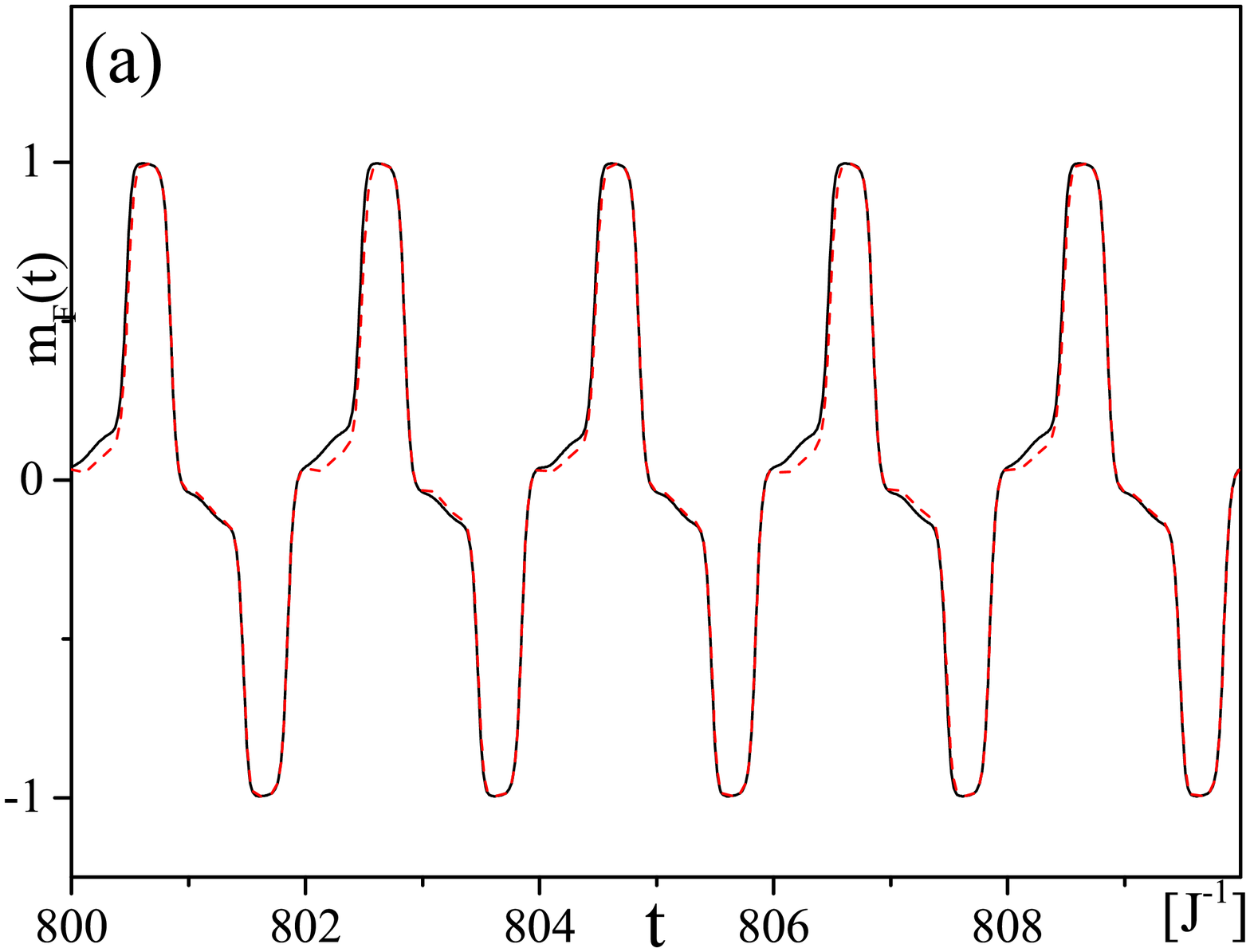}
\includegraphics[width=0.9\linewidth,bb=80 50 770 545]{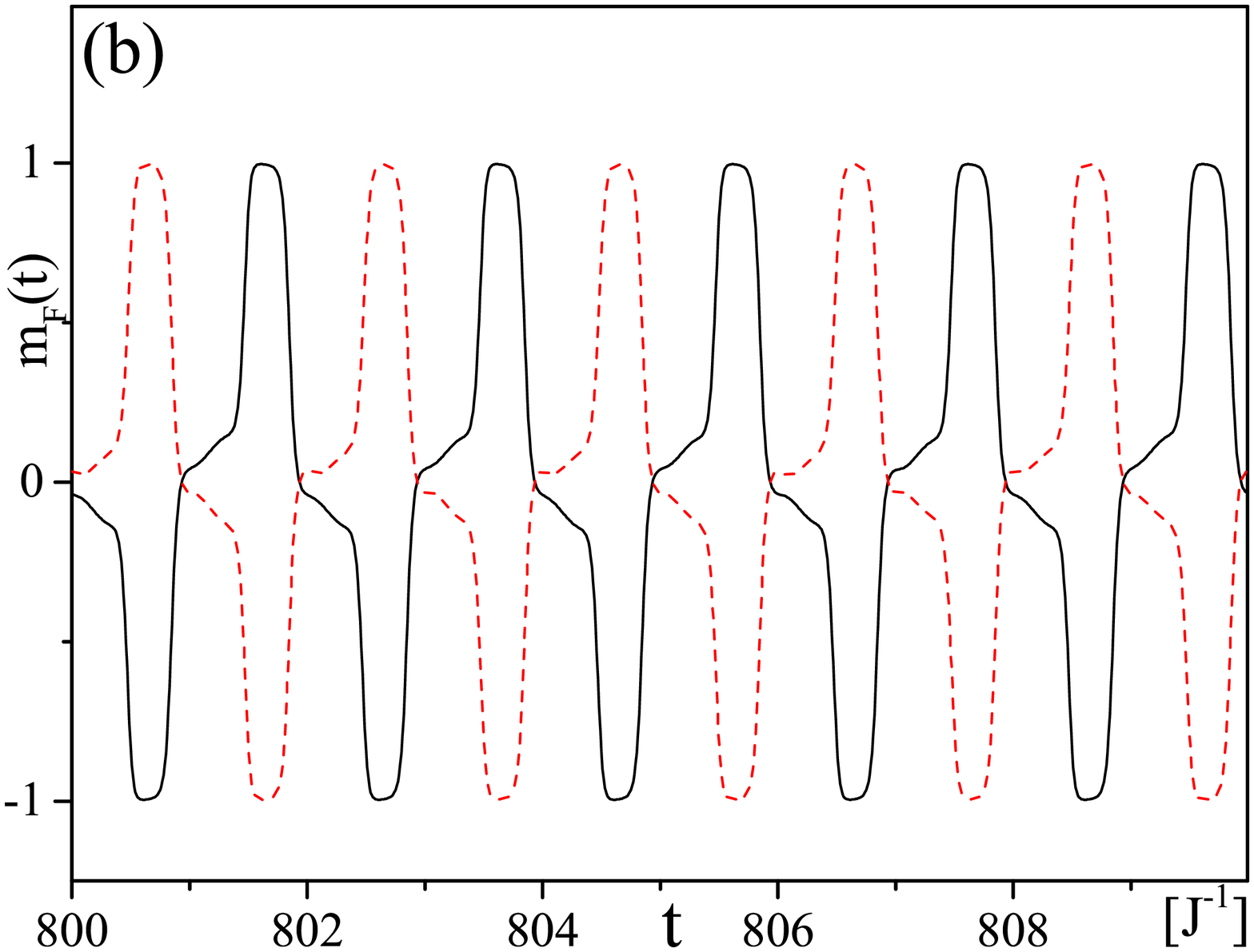}
\caption{(Color online) Comparison between the dynamics of FM order parameter $m_F(t)$ with (a) different noise trajectories and (b) different initial states with $J'=5J$ and $L=30$. Other parameters are chosen the same as in Fig.\ref{fig:SM2}. } \label{fig:SM3}
\end{figure}

\subsection{ Convergence of numerical results}
{\it Finite $\Delta$t --} Throughout the maintext, we choose a discrete time step of $\Delta t=10^{-3}J^{-1}$.  For a stochastic differential equation,  the choice of  $\Delta t$ is more subtle than that in the deterministic equation since the random variable depends on $\Delta t$ as shown in Eq.(\ref{eq:random}). To examine the convergence of our result with $\Delta t$, we choose different $\Delta t J=2\times 10^{-3}, 10^{-3}$ and $5\times10^{-4}$, and compare their results. As shown in Fig.\ref{fig:SM2} (a), the results with different  $\Delta t$  agree with each other very well, which indicates that the $\Delta t$ chosen in our simulation is sufficiently small, thus enables us to ignore the errors induced by the discretization of time.

{\it Finite L --} In the maintext,  the system we simulated is up to a system size with $L=120$. One needs to check the system size dependence of our results. For an ordered phase ({\it e.g.}, the DTC phase with strong driving ($J'=5J$)),  as shown in Fig.\ref{fig:SM2} (b), the deviation between the  results with $L=30$, $60$ and $120$ is pretty small.  However, in the intermediate driving regime without true long-range ferromagnetic (FM) order, the FM order parameter $m_F(t)$ strongly depends on the system size. As shown in Fig.\ref{fig:SM2} (c),  the amplitude of $m_F(t)$ significantly decreases with $L$.  This is due to the fact that in the presence of intermediate driving,  the long-range FM order has not been built up during the FM coupling. Instead, the system is spontaneously separated into different FM domains, and the overall FM order parameter is a summation of the magnetization of them.  Within each FM domain, the magnetization oscillates as a DTC, but the phases of these DTCs are not coherent, thus the magnetization in different domains at a given time cancels with each others. As a consequence, the overall FM order parameter decreases with the system size.

The different finite size dependence of the FM order parameter between the strong and intermediate driving can be considered as the signature of the different dynamical phases with long-range and short-range FM orders respectively. In addition, the finite size effect is supposed to be important near the dynamical critical points, which is an important issue but  not addressed in this work.

{\it Noise trajectories --} In principle, one needs to simulate over different noise trajectories and perform the ensemble average over them. However, in our simulation, we only randomly choose one noise trajectory for each set of parameters. This is because we  are only interested in the situation with  weak noise($\mathcal{D}=0.01J$), where a small thermal fluctuation does not change the nature of the phases with discrete symmetry breaking (see the comparison between the dynamics over two different noise trajectories in Fig.\ref{fig:SM3} a). However, for some special initial states, it is possible that the system could be trapped into a metastable state if there is no thermal fluctuations. The role of noise in our simulation is to thermally activate the system and make it escape from the metastable state after sufficiently long time and enter the genuine asymptotic long-time states discussed in the maintext.

{\it Initial states --} In our simulation, we start from a spatially inhomogeneous random initial state: for each site, we choose its  initial state as $\bm{s}_i^0=[s_i^x,0,s_i^z]$, where $s_i^x$ is an random number different from site by site and uniformly distributed within [-1,1], the z-component of the spin is chosen  correspondingly as $s_i^z=\sqrt{1-[s_i^x]^2}$. Since we focus on the non-equilibrium phases with spontaneously symmetry breaking, it is well known that in this case, the final state is supposed to be highly sensitive to the initial state. This statement does not only work for equilibrium phases (FM or AF order), but also for non-equilibrium phases. For instance, for a DTC phase with spontaneously $Z_2$ time translational symmetry breaking, as show in Fig.\ref{fig:SM3} (b), starting from different initial states, the system could finally fall into either one of the $Z_2$ breaking phases, each of which is related with the other by a half-period shift in time domain.

 \begin{figure}[htb]
\includegraphics[width=0.9\linewidth,bb=80 50 770 545]{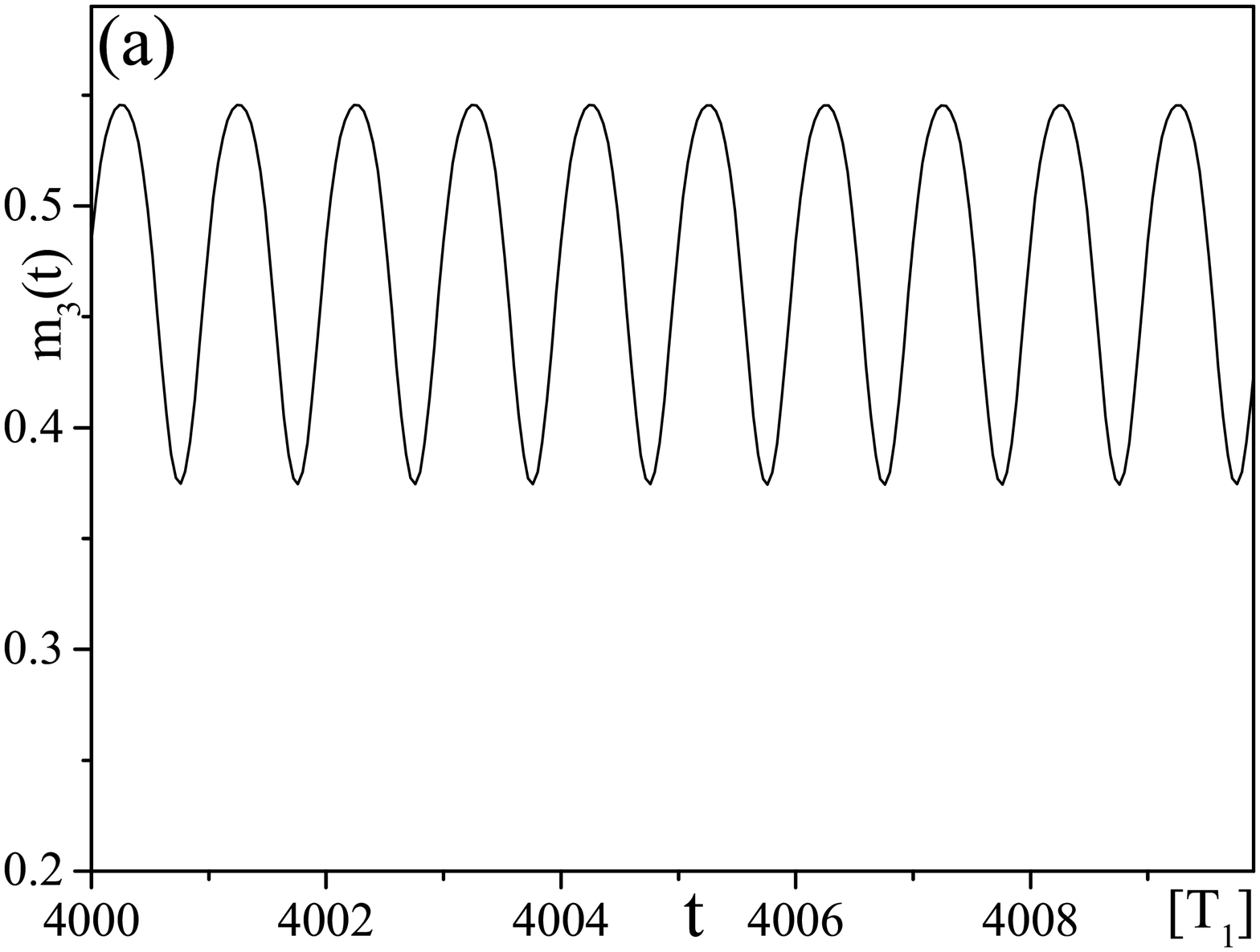}
\includegraphics[width=0.9\linewidth,bb=80 50 770 545]{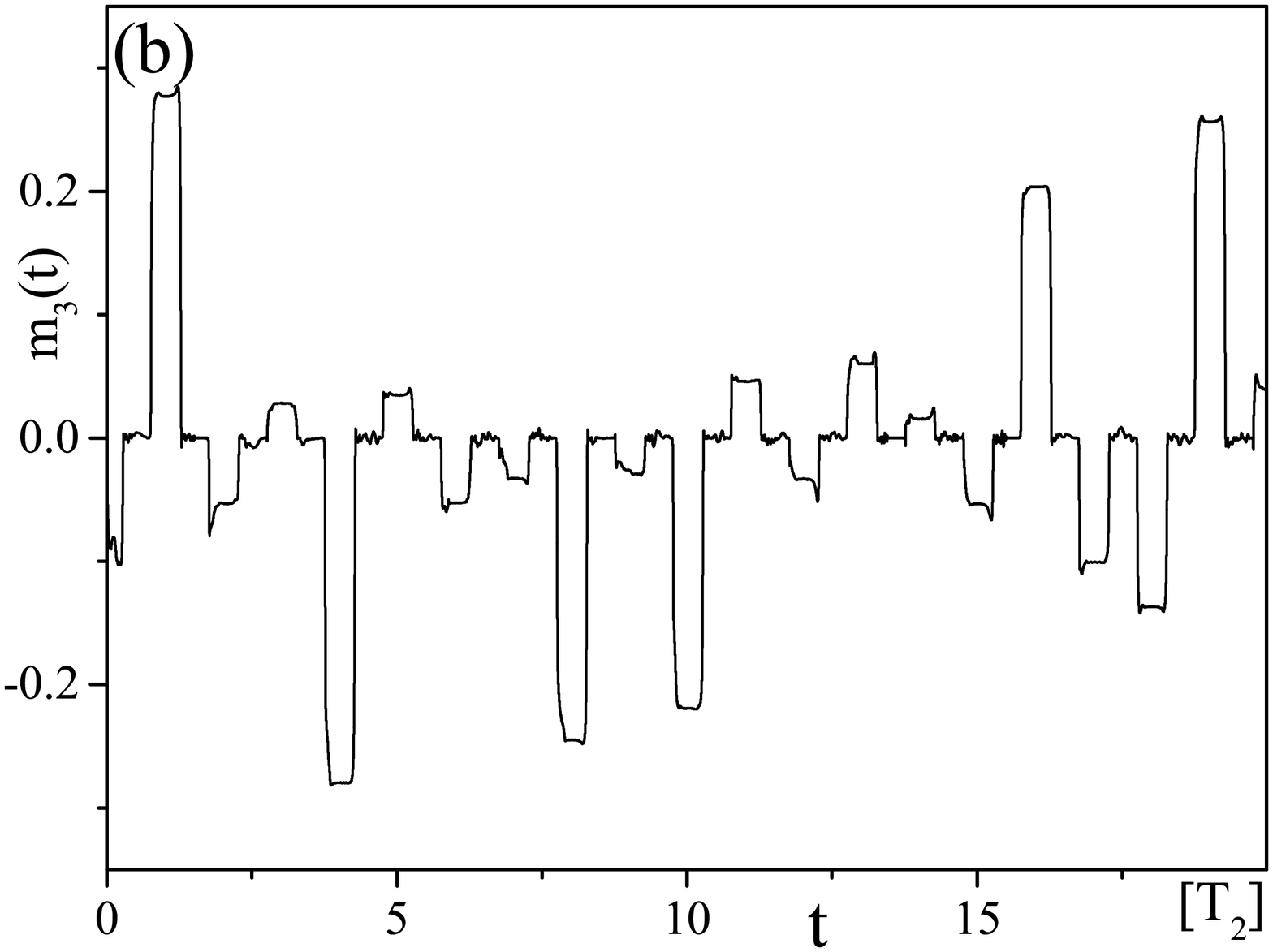}
\caption{(Color online) The dynamics of the 3-period stripe order parameter $m_3(t)$ in the presence of (a) fast driving with a frequency $\omega=8 \pi J$ and (b) slow driving with a frequency $\omega=0.02\pi$.  The x-axis are plotted in the unit of the corresponding driving periods with $T_1=0.25J^{-1}$ and $T_2=100J^{-1}$. The parameters are chosen as $L=30$, $J'=5J$. Other parameters except $\omega$ are chosen the same as in Fig.\ref{fig:SM2}. } \label{fig:SM4}
\end{figure}

\section{The role of different parameters in the model}
In the maintext, we only studied several representative non-equilibrium phases by focusing on special points in the parameter space.  In the following, we will systematically examine the role of different parameters of the models in determining the space-time patterns of dynamical phases. The dynamical phase transitions between them have also been studied.

 \begin{figure*}[htb]
\includegraphics[width=0.325\linewidth,bb=80 50 770 545]{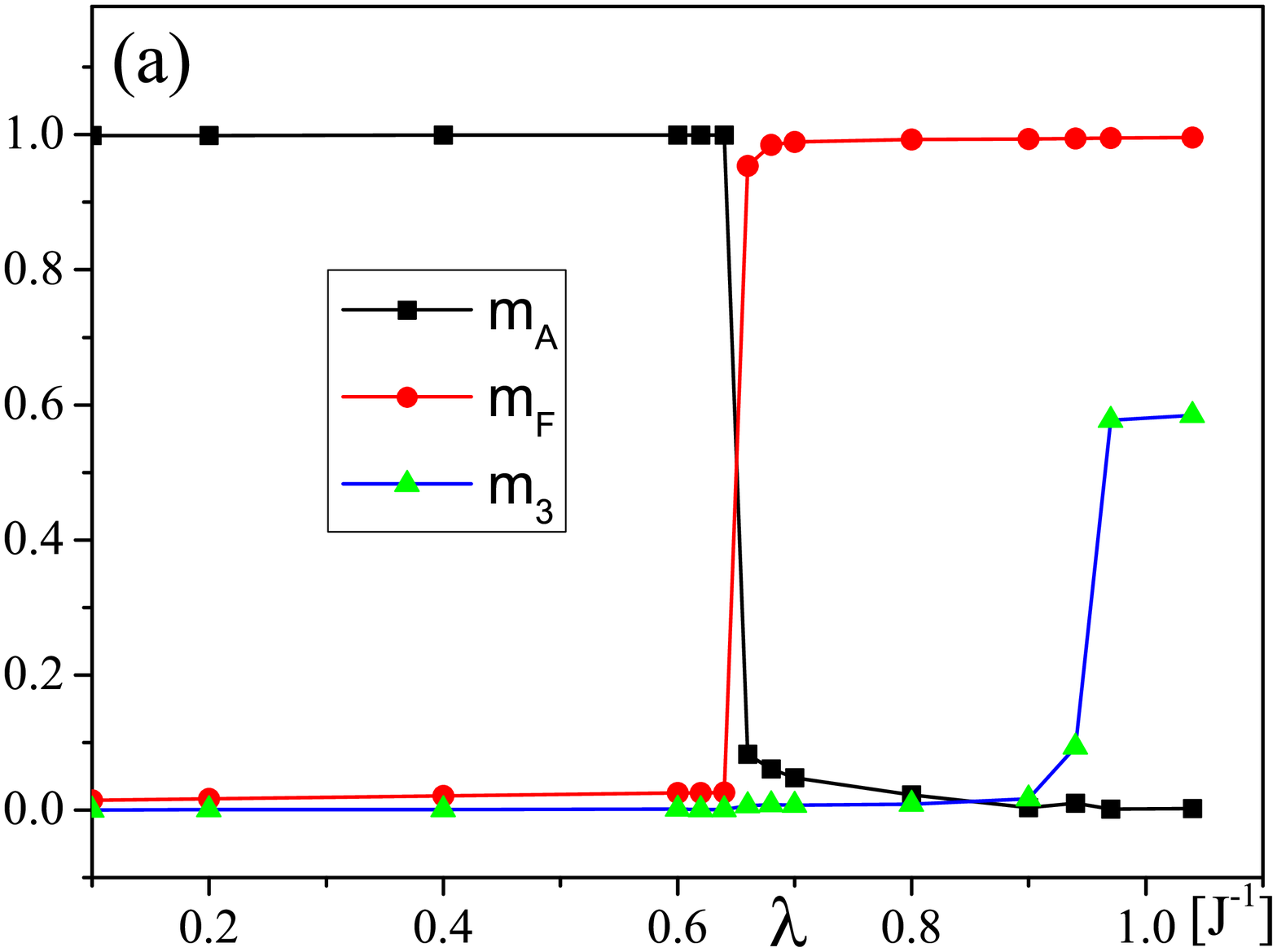}
\includegraphics[width=0.325\linewidth,bb=80 50 770 545]{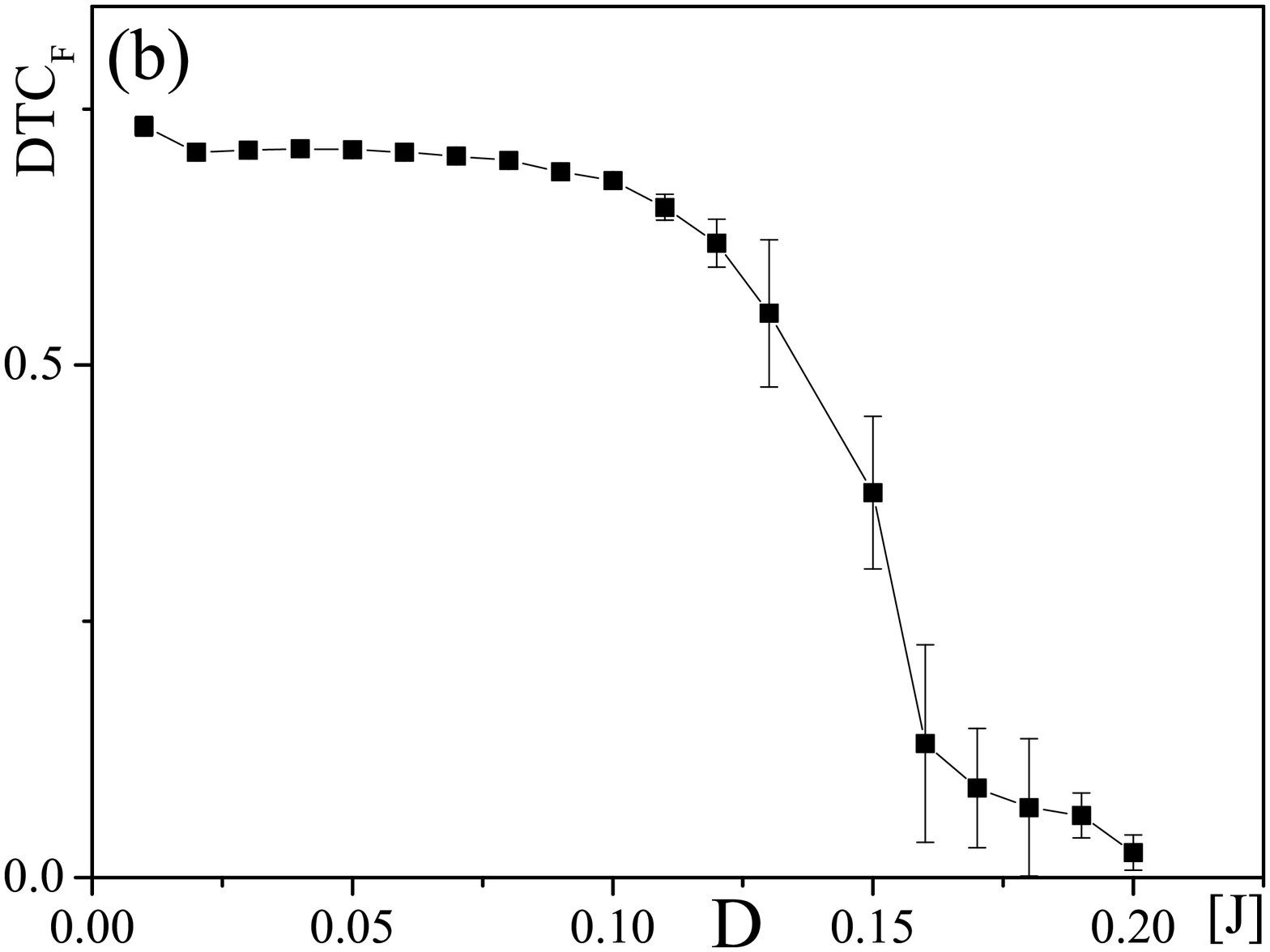}
\includegraphics[width=0.325\linewidth,bb=80 50 770 545]{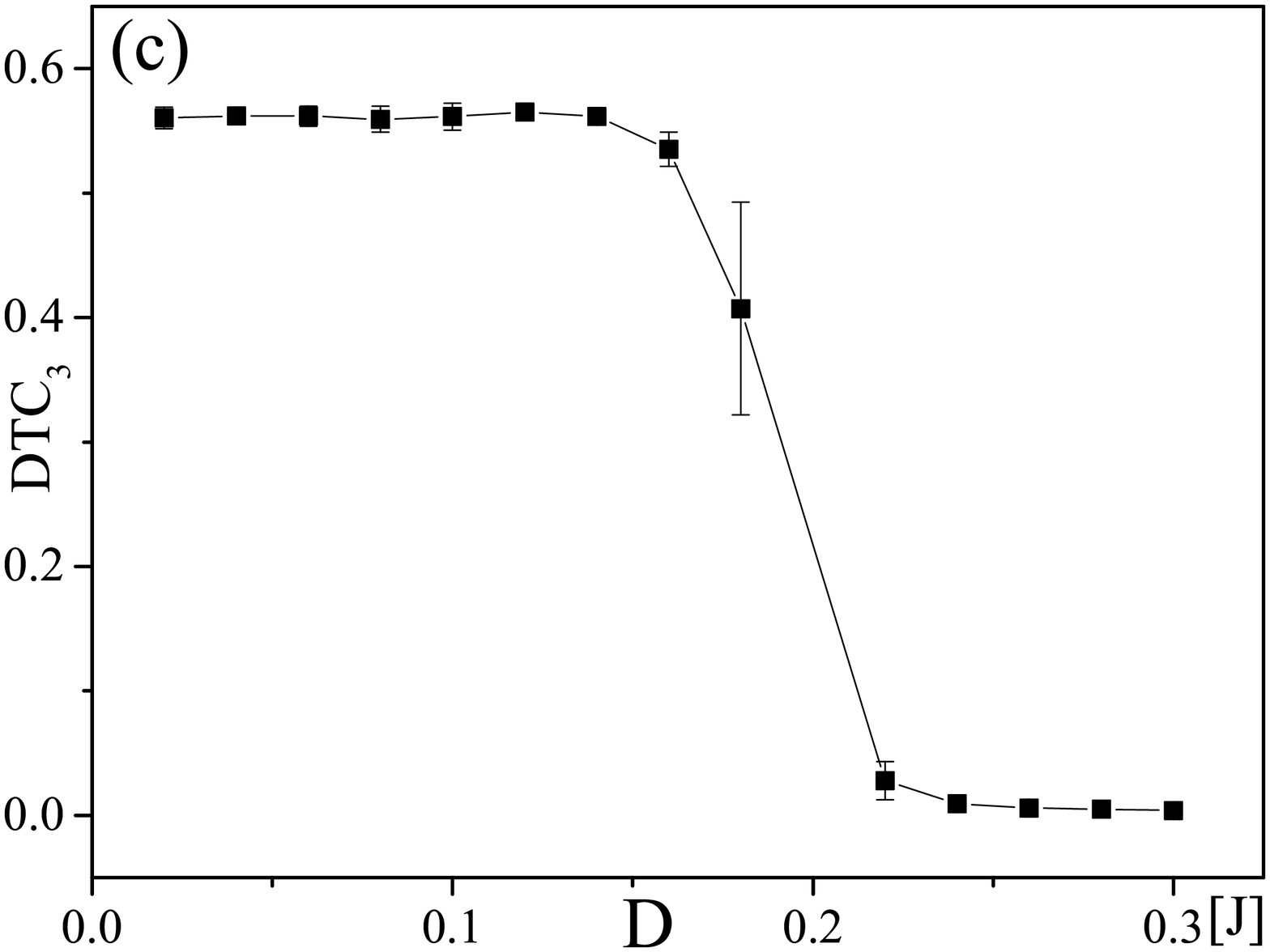}
\caption{(Color online) (a) The time-independent AF ($m_A$), FM ($m_F$) and tripartite stripe ($m_3$) order parameters as a function of frustration strength $\lambda$  with $J'=5J$ and $\mathcal{D}=0.01J$. (b) In the strongly driven case ($J'=5J$), the DTC order parameter corresponding to the FM order as a function of noise strength $\mathcal{D}$.  (c)In the intermediate driven case ($J'=3J$), the DTC order parameter corresponding to the 3-period stripe order as a function of noise strength $\mathcal{D}$. $\lambda=J$ for (b) and (c). Other parameters are chosen as $L=30$,  $\eta=J$, $h_z=1.5J$, $\omega=2\pi$. } \label{fig:SM5}
\end{figure*}

\subsection{Drving frequency $\omega$}
In the maintext, we changed the driving amplitude but fixed the driving frequency as $\omega=2\pi J$. Here, we will exam the role of $\omega$ in determining the space-time patterns of our model.

In the fast driving limit where $\omega\gg J$, the periodic driving oscillates too fast to be followed by the system. In such a high frequency limit, one can derive an effective time-independent Hamiltonian to describe the stroboscopic dynamics of this periodically driven system, similar as the Floquett analysis in quantum systems, where the effective time-dependent Hamiltonian can be expressed as a expansion in terms of $1/\omega$. In the high frequency limit, the dominant term in the expansion is an average of the Hamiltonian over one period, where $\frac 1{T_0}\int_0^{T_0} dt V(t)=J$, therefore, the dynamics in this case is similar with the relaxation dynamics in the undriven case ($J'=0$), where the steady state is a stripe phase with a nonvanishing order parameter $m_3=\frac 1{L^2}\sum_\mathbf{i} \sin[\mathbf{Q}_0\cdot\mathbf{i}]s_\mathbf{i}^x(t)$ with $\mathbf{Q}_0=(\frac {2\pi}3,\frac{2\pi}3)$. For a large but finite frequency, the order parameter of the stripe phase will oscillates around its equilibrium value with a frequency the same as the driving,  as shown in Fig.\ref{fig:SM4} (a) where the frequency is chosen as $\omega=8\pi J$.

In the opposite limit of slow driving, where the period of the driving is much longer than the relaxation time $\omega\ll \eta$ ($\eta$ is the dissipation strength), the system have sufficient time to relax, thus at any given time, the system is always close to an equilibrium state. As a consequence, both the FM and the stripe spatial order can be developed depending on the sign of $V(t)$ in the instantaneous Hamiltonian. However, a thermalization of a system means that it has no memory of the information of its initial state, or the previous states far away from it. As a consequence, for magnetic states with SSB, the system will randomly choose one state among the degenerate manifold with SSB, while these symmetry breaking phases at different time slices barely correlate with each other, thus cannot form a long-range order (DTC) in time domain, as shown in Fig.\ref{fig:SM4} (b) where the frequency is chosen as $\omega=0.02\pi$.

\subsection{Frustration $\lambda$}

In the maintext, we only studied the unfrustrated ($\lambda=0$) and full-frustrated ($\lambda=1$) cases, which exhibits non-equilibrium phases with different SSB. Here, we will systematically study the role of frustration by continuously tuning the frustration strength $\lambda$.  We focus on  strongly driving case ($J'=5J$). We find that the frustration does not change the DTC nature of the phase, but is crucial in determining the spatial order of the non-equilibrium phases. In general, the magnetic order parameters in these dynamics phases keep oscillating in time, therefore, to characterize their strength, we need to derive a time-independent order parameter.   For instance, for the FM order parameter $m_F(t)$, we choose those time slices with $t=t_n^F$ when $m_F(t)$ reach its n-$th$ maximum, and perform the average over them to derive a time-independent order parameter $m_F=\langle m_3(t_n^F)\rangle$, and use it to characterize the strength of the FM order parameter in these dynamic phases.

We plot the time-independent AF, FM and tripartite stripe order parameters  $m_A$, $m_F$ and $m_3$ as a function of $\lambda$ in Fig.\ref{fig:SM5} (a), from which we can find that for a small frustration, the AF-DTC order persist until $\lambda=0.68J$, where the AF order gives way to a FM order via a first order (1st) phase transition. For a frustration within the regime $\lambda \in [0.68J,0.9J]$, one can find  an intermediate phase where  the long-range FM order has been built up in the duration of FM coupling, while there is no long-range tripartite stripe order during the AF coupling. In another word, frustration suppresses the AF order even in this non-equilibrium driven case, while it facilitates the FM orders, since in the duration with FM coupling ($V(t)<0$), the next-nearest-neighboring coupling no longer leads to ``frustration'', instead it increases the effective FM coupling.    When the frustration further increases, the system experience another 1st order  dynamical phase transition at $\lambda=0.9J$ characterized by the sudden onset of the tripartite stripe order, and the system enters a dynamics phase oscillating between the states with long-range FM and tripartite stripe orders, which has been discussed in maintext of the fully-frustrated strongly driving case.

\subsection{Thermal fluctuation $\mathcal{D}$}
In the maintext, we focus on the case with weak thermal fluctuation ($\mathcal{D}=0.01J$), which does not change the nature of the phases with discrete symmetry breaking.  However, it is known that the thermal fluctuation works against spontaneous symmetry breaking, and a strong thermal fluctuation could melt the ordered phase and restore the symmetries. Since the space-time crystals phase discussed here simultaneously breaks different symmetries (space and time translational symmetries and $Z_2$ spin symmetry), one may wonder what's the effect of the thermal fluctuations on these dynamical orders.

As an example, we  focus on the DTC order with spontaneous $Z_2$ symmetry breaking in time domain,  and check whether it is possible for thermal fluctuation to restore this symmetry. Since at low temperature, both $m_F(t)$ and $m_3(t)$ exhibit DTC order in time domain, one needs to distinguish their corresponding DTC order via different order parameters as:
\begin{equation}
DTC_a= \frac 2{t_0}\int_{\frac{t_0}2}^{t_0}dt e^{i\pi t} m_a(t)
\end{equation}
with $a=F$ or $3$  indicates the $FM$ or tripartite stripe order parameter respectively.  $t_0=2000J^{-1}$ is our simulation time and the Fourier transformation is performed over the second half of the full simulation time, during which the system has reached the long-time asymptotic state.

We focus on both cases with intermediate ($J'=3J$) and  strong ($J'=5J$) driving  and study the corresponding DTC order parameters  $DTC_3$ and $DTC_F$ as a function of $\mathcal{D}$. The ensemble average is performed over $\mathcal{N}$ noise trajectories, with $\mathcal{N}=10$. For the case with strong driving, the DTC order parameter corresponding to the FM order persist until $\mathcal{D}\approx 0.15J$, above which the discrete time translational symmetry has been restored (Fig. \ref{fig:SM5} b), however, the DTC order corresponding to the tripartite stripe order is much more fragile, it vanishes for $\mathcal{D}>0.03J$ (not shown here). In summary, in the case with strong driving, there exists an intermediate noise regime where the FM-DTC order survives but the stripe-DTC order does not, similar with what happened in the intermediate frustration regime discussed above. In the case with intermediate driving, the long-range FM order has not been built up even in the presence of weak noise and only tripartite stripe order exists.  However, different from its counterpart in the strongly driven case, the DTC order corresponding to this tripartite stripe phase in the intermediately driven case is quite robust against thermal fluctuation, as shown in Fig. \ref{fig:SM5} (c).  The thermal fluctuation-induced transitions discussed in these two cases seem to be continuous. However, a precise determination of the position of the phase transition point and the critical properties calls for a finite-size scaling analysis, which will be left in the future work.


\end{document}